\documentclass[conference]{IEEEtran}

\usepackage[english]{babel}
\usepackage{blindtext}
\usepackage{cite}
\usepackage{amsmath,amssymb,amsfonts}
\usepackage{algorithmic}
\usepackage{graphicx}
\usepackage{textcomp}
\usepackage{xcolor}

\usepackage{hyperref}
\hypersetup{
hidelinks,
colorlinks=true,
linkcolor=red,
citecolor=blue,
anchorcolor=red,
breaklinks=false,
}

\usepackage{enumitem}
\usepackage{subfig}

\usepackage{caption}
\usepackage{graphicx}
\usepackage{float}
\usepackage{array}
\usepackage{booktabs}
\definecolor{azure_blue}{rgb}{0.0, 0.5, 1.0}
\usepackage{tikz}

\usepackage{amsmath}

\newcommand{\squishlist}{
  \begin{list}{$\bullet$}{
    \setlength{\itemsep}{0pt}       \setlength{\parsep}{3pt}
    \setlength{\topsep}{3pt}        \setlength{\partopsep}{0pt}
    \setlength{\leftmargin}{1em}    \setlength{\labelwidth}{1em}
    \setlength{\labelsep}{0.5em} } }

\newcommand{\squishend}{
  \end{list} }

\newcommand{\zheng}[1]{#1}
\newcommand{\sideword}[1]{}

\begin{document}
\begin{sloppypar}
    
\date{}

\title{Demystifying Datapath Accelerator Enhanced Off-path SmartNIC}

\author{
    \IEEEauthorblockN{Xuzheng Chen$^{\dag,1}$, Jie Zhang$^{\dag,1}$, Ting Fu$^2$, Yifan Shen$^2$, Shu Ma$^2$, Kun Qian$^2$, \\Lingjun Zhu$^2$, Chao Shi$^2$, Yin Zhang$^1$, Ming Liu$^3$, and Zeke Wang$^1$}
    \IEEEauthorblockA{$^1$ Zhejiang University\qquad $^2$ Alibaba Cloud\qquad $^3$ University of Wisconsin–Madison}
}

\IEEEoverridecommandlockouts\IEEEpubid{\makebox[\columnwidth]{ 979-8-3503-5171-2/24/\$31.00 $\copyright$2024 IEEE \hfill}\hspace{\columnsep}\makebox[\columnwidth]{ }}

\maketitle
\begin{abstract}
    Network speeds grow quickly in the modern cloud, so SmartNICs are introduced to offload network processing tasks, even application logic. However, typical multicore SmartNICs such as BlueFiled-2 are only capable of processing control-plane tasks with their embedded processors that have limited memory bandwidth and computing power. On the other hand, cloud applications evolve rapidly, such that a limited number of fixed hardware engines in a SmartNIC cannot satisfy the requirements of cloud applications. Therefore, SmartNIC programmers call for a programmable datapath accelerator (DPA) to process network traffic at line rate. However, no existing work has unveiled the performance characteristics of the existing DPA. 

    To this end, we present the first architectural characterization of the latest DPA-enhanced BlueFiled-3 (BF3) SmartNIC. Our evaluation results indicate that BF3's DPA is significantly wimpier than the off-path Arm processor and the host CPU. However, we still identify that DPA has three unique architectural characteristics that unleash the performance potential of DPA. Specifically, we demonstrate how to take advantage of DPA's three architectural characteristics regarding computing, networking, and memory subsystems. Then we propose three important guidelines for programmers to fully unleash the potential of DPA. To demonstrate the effectiveness of our approach, we conduct detailed case studies regarding each guideline. Our case study on key-value aggregation achieves up to 4.3$\times$ higher throughput by using our guidelines to optimize memory combinations.
\end{abstract}

\vspace{-1ex}
\section{Introduction}
{
\renewcommand{\thefootnote}{}
\setlength{\skip\footins}{0cm}
\footnotetext{$^{\dag}$Equal contribution}
}
In the modern cloud, network speeds grow quickly, with 100/200 Gigabit Ethernet (GbE) network interface controllers (NICs) widely deployed~\cite{broadcom_100G,mellanox_cx5,mellanox_cx6} and 400/800 GbE expected in the near future~\cite{broadcom_400G,eth_800G}. At the same time, Moore’s law is slowing down, and the gap between network and CPU speeds is rapidly increasing. More and more works intend to equip NICs with more computing power to alleviate the computing pressure of host servers.

These NICs with computing power are usually called SmartNICs, e.g., multicore system-on-chip~(SoC) SmartNICs. SoC SmartNICs provide significantly more generality and have been widely used in various application domains~\cite{bf2}. 
Prior work~\cite{ipipe} categorizes SoC SmartNIC into two types: on-path and off-path, according to how SmartNIC processors interact with network traffic. Processors of on-path SmartNICs sit on the packet path and can directly manipulate each incoming/outgoing packet. For off-path SmartNICs, incoming packets from the network are delivered to either host CPUs or off-path processors based on forwarding rules installed on the NIC switch, while both host CPUs and off-path processors can send out packets through the NIC TX port. 

The processors in the SoC SmartNICs are usually Arm~\cite{bf2,bf3,broadcom_stingray_ps250} or cnMIPS~\cite{liquidio2}, and have fewer threads compared with the host CPU. This prevents the SmartNIC processors from processing network traffic at line rate. Current off-the-shelf SmartNICs solve this problem by equipping NICs with a few powerful hardware domain-specific engines, such as erasure coding, de/compression, and encryption engines~\cite{bf2,bf3,ipu,liquidio2,broadcom_stingray_ps250}. However, different cloud applications have different offloaded computation kernels, and fifty hot cloud applications occupy around 60\% host CPU cycles~\cite{profiling_isca15, challenge_in_cloud, ndp}, each hot application could need a customized engine to accelerate~\cite{profiling_isca23, tiara}, and thus Google even proposes the concept of “a sea of accelerators” for cloud applications. Even worse, hot cloud applications change over time~\cite{profiling_isca15,beyond_smartnic, transport_protocol_in_nic}. 
Therefore, the current SoC SmartNICs that feature a few fixed hardware engines fail to meet the requirements of cloud applications. 

This trend calls for a programmable datapath accelerator that can provide line-rate data processing ability while serving a broad range of applications. SmartNIC vendors are trying to integrate a programmable datapath accelerator in the SmartNIC datapath. Nvidia's latest BlueField-3 (BF3)~\cite{bf3} adds a many-core RISC-V processor in the datapath (called datapath accelerator, DPA). Even though plenty of recent works~\cite{battle_bf3_hoti,bf2_compression_bench,dpubench,fudan_bf2,ipads_bf2,ipipe,bf2_performance, understanding_lumina} have characterized the traditional SmartNICs, there is no comprehensive study on DPA-enhanced SmartNICs. Therefore, it's still unclear for programmers to fully understand the performance characteristics of DPA, which could heavily interact with off-path processors and host CPUs.




To this end, this paper conducts the first systematic benchmark on characterizing the performance of DPA-enhanced SmartNIC, specifically the BlueField-3. We thoroughly evaluate the resources in a BF3-attached server including general-purpose computing power. To tap into the potential of DPA, we assess it from an architectural perspective instead of reporting performance numbers. Our evaluation results indicate that the current DPA is markedly underperforming than the host/Arm. However, we still identify that DPA has three unique architectural characteristics regarding computing, networking, and memory subsystems, that expose the potential to benefit certain types of applications. We demonstrate how to take advantage of DPA's three architectural characteristics to fully unleash the potential of the underperforming DPA by using three corresponding case studies. The experimental results show that the achievable throughput of the key-value aggregation service can be increased by up to 4.3$\times$ and the time uncertainty bound of the clock synchronization service can be decreased by up to 2.3$\times$. Then we conclude three important DPA-related guidelines for future SmartNIC programmers regarding these architectural characteristics: 

\squishlist

\item \textbf{1. Offloading latency-sensitive and simple workloads.}
Compared with the host/Arm, DPA is much closer to the network, since DPA and NIC are on the same chip. As such, DPA enjoys the lowest network latency. Latency-sensitive network applications can exploit this characteristic to improve end-to-end performance. Our case study on clock synchronization service achieves up to 2.3$\times$ lower time uncertainty bound.\sideword{R1} \zheng{However, we cannot offload compute- and/or memory-bandwidth-bound logic to DPA due to its limited computing ability and memory bandwidth of a single DPA thread.} 

\item \textbf{2. Offloading easy-to-parallelize workloads with small working set sizes.}
\sideword{R1}\zheng{Our benchmarking results reveal that offloading stateless network functions to DPA cores can achieve line rate as host/ARM, even though each DPA thread is significantly wimpier, because DPA has more cores than the host/Arm.} However, when the application's working set size exceeds DPA's cache size, significant performance downgradation would occur. Programmers can offload easy-to-parallelize workloads to exploit DPA's many-core parallelism. Also, the memory working set size should better fit in DPA's cache size to avoid significant degradation. 

\item \textbf{3. Carefully select memory buffers when running network workloads in DPA cores.}
DPA can access not only its own memory but also the off-path Arm's memory and the host CPU's memory. Using different memories for the DPA cores can result in several times performance differences. Programmers must carefully choose different memories according to the specific usage. Our case study on key-value aggregation service achieves up to 4.3$\times$ higher throughput by optimizing memory combinations. 

\squishend

\sideword{R2}\zheng{We acknowledge that our findings may not fully apply to the next generation of BlueField SmartNICs or products from other vendors. However, we believe that our methodology (i.e., studying the characteristics of each component and offloading workloads accordingly) can be applied to other SmartNICs. Our benchmark code and tools are available at \urlstyle{sf}\url{https://github.com/RC4ML/BenchBF3}.}


\vspace{-1ex}
\section{Background}
\subsection{On-path/Off-path SmartNIC}
SoC SmartNICs comprise a multicore processor (i.e., MIPS/ARM) and the processor is usually underperforming due to the cost, form factor, and power~\cite{rambda_hpca23}. SoC SmartNIC has its own onboard SRAM/DRAM and usually has a DMA engine to access host server memory. A different SoC SmartNIC has a different set of domain-specific accelerators (e.g., encryption, regular expression matching, and erasure coding). 

\begin{figure}
	\centering
	{\includegraphics[keepaspectratio=true, width=0.9\linewidth]{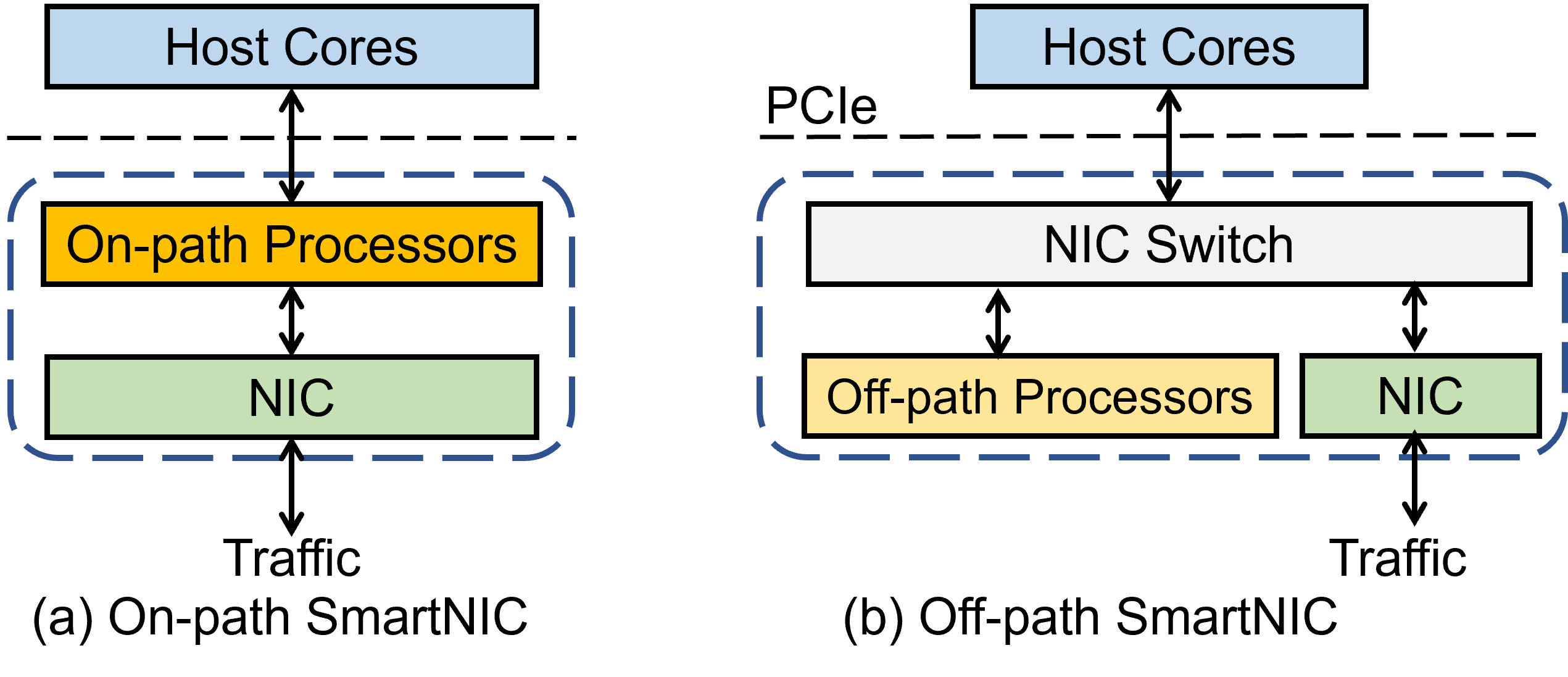}
    \vspace{-1ex}

	\caption{On-path and Off-path SmartNICs.}
	\label{fig_on_off_path}}
    \vspace{-2ex}
\end{figure}

Depending on where the processor sits on the network path, SmartNICs can be categorized into two categories: On-path and Off-path. As shown in Figure~\ref{fig_on_off_path}a processors in on-path SmartNICs provide inline processing power for each incoming/outgoing packet. Figure~\ref{fig_on_off_path}b shows the schematic diagram of off-path SmartNICs, and each incoming packet can be forwarded to SmartNICs processor or host CPU. This kind of pattern is also called lookaside acceleration~\cite{k_lookaside}. Processors of on-path SmartNICs are usually thought to be closer to the network than off-path SmartNICs due to the higher wire latency introduced by the NIC switch ($\sim$500 ns in BF3).

\begin{figure}
	\centering
	{\includegraphics[keepaspectratio=true, width=0.9\linewidth]{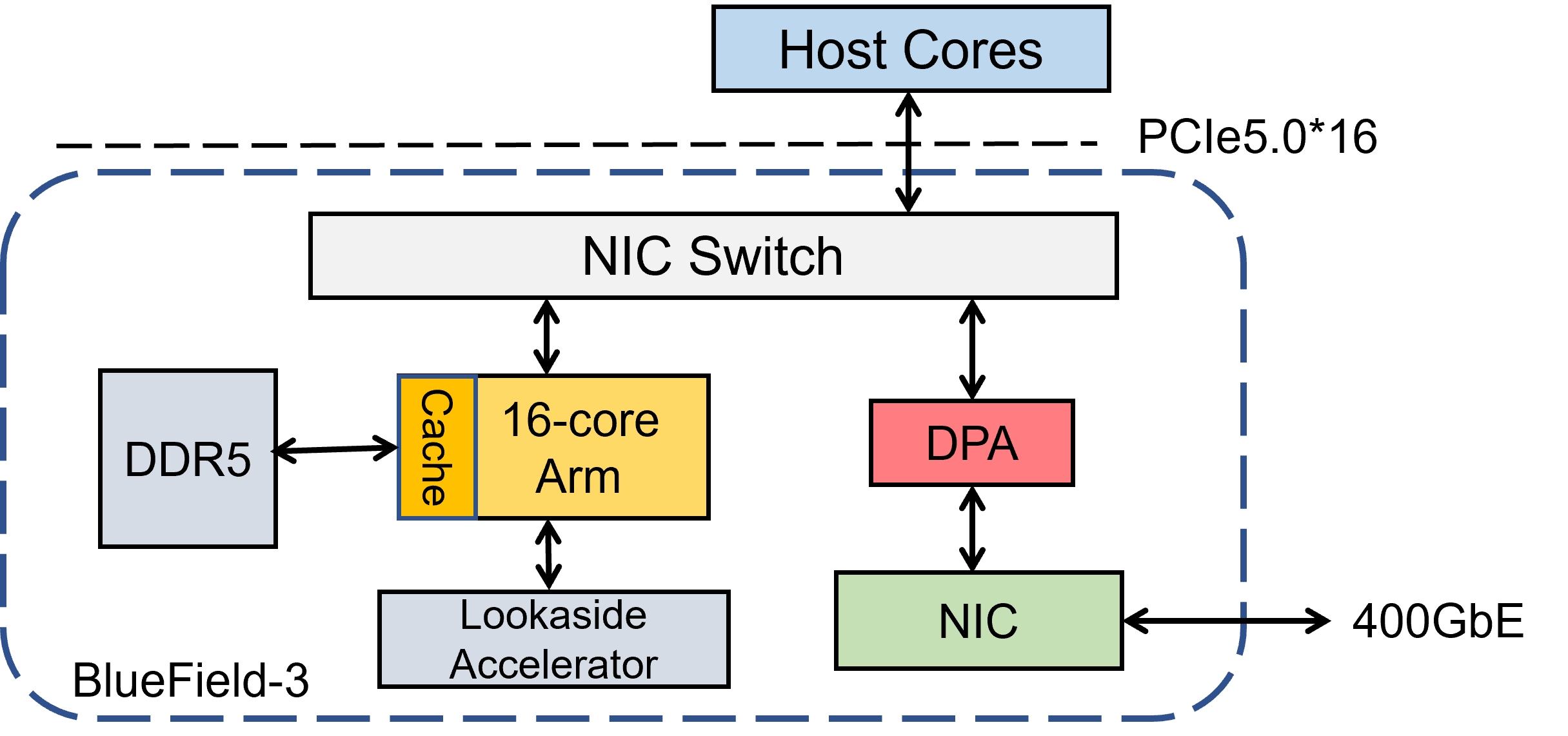}

	\caption{BlueField-3 SmartNIC architecture.}
	\label{fig_bf3_arch}}
    \vspace{-3ex}
\end{figure}

\subsection{\bf DPA-enhanced BlueField-3 SmartNIC }
Figure~\ref{fig_bf3_arch} demonstrates the overall architecture of BF3 SmartNIC. BF3 mainly consists of an off-path Arm processor, a many-core datapath RISC-V processor, an off-chip DDR5 memory, a PCIe switch, several domain-specific accelerators, and a NIC module. 

Unlike the off-path Arm processor and domain-specific accelerators that provide look-aside computing power, the many-core DPA sits in the network critical path, providing inline processing ability. Unlike a naive on-path processor, DPA is designed to be able to tightly interact with the Arm-related resources and the host-related resources. Despite DPA's three-level cache and memory, DPA can access the host's last-level cache (LLC), host memory, Arm LLC, and Arm memory. The complicated cache/memory hierarchy makes it challenging for programmers to fully unlock the potential of the datapath accelerator-enhanced off-path SmartNIC. 

\subsection{Experiment Setup}
\label{sec_server_setup}

We build a testbed and conduct various experiments on it to fully understand the performance characteristics of the abundant resources of DPA-enhanced BF3. \sideword{R3}\zheng{The testbed comprises two servers running Ubuntu 22.04 (Linux kernel-5.15.102), each equipped with a BF3 SmartNIC.} The two BF3 SmartNICs are connected back-to-back using two 200GbE QSFP56 cables. The detailed configuration of the servers is shown in Table~\ref{tab_experiment_setup}. In all network-related experiments, we use link aggregation~\cite{link_aggregation} to combine two network interfaces into a single interface and use hash mode to distribute packets. DPA software uses NVIDIA DOCA framework~\cite{doca} (v2.5.0) to process network packets. Currently, only 190 out of 256 DPA threads can be used concurrently due to the limitation of the DOCA driver. As such, we use at most 190 DPA threads in related experiments. \sideword{R3}\zheng{The host/Arm software leverages DPDK~\cite{dpdk} (v22.11) to process network packets and uses RSS~\cite{rss} to distribute packets among different queues (one queue per core) unless stated otherwise.}

The detailed comparison of three kinds of compute resources is shown in Table~\ref{tab_compute_resources}. The noticeable difference is that DPA has markedly more threads than the host/Arm, so it becomes necessary to benchmark the DPA-enhanced SmartNIC such that programmers can easily optimize their applications on top of our benchmarking hints.  

\begin{table}[]
    \centering
    \vspace{2ex}
    \setlength{\extrarowheight}{1pt}
    \caption{Hardware description of experiment setup}
    \vspace{-1ex}
    \begin{tabular*}{\linewidth}{l l}
    \toprule   \textbf{Component} & \textbf{Hardware description} \\   
    \hline   Server & SuperServer SYS-421GE-TNRT  \\ 
       Host CPU &  Intel Xeon Gold 6426Y   \\  
       Host Memory & 256 GB DDR5-4800 (8 out 8 channels)  \\
       NIC   &  1$\times$ BlueField-3 B3220(2 $\times$ 200Gbps) \\
       NIC PCIe   & PCIe 5.0 $\times$16 \\
    \bottomrule   
    \end{tabular*}   
    \label{tab_experiment_setup}
    \vspace{-3ex}
\end{table}

\begin{table}[]
    \centering
    \vspace{2ex}
    \setlength{\extrarowheight}{1pt}
    \caption{Hardware Specifications of compute resources}
    \vspace{-1ex}
    \scalebox{1}{
        \begin{tabular}{l l l l}
        \toprule   \textbf{Resources} & \textbf{Host (X86)}  & \textbf{Arm} & \textbf{DPA (RISC-V)} \\   
        \hline   Processor & Intel Xeon & Cortex-A78AE & RV64IMAC  \\ 
           Cores & 16 & 16 & 16 \\
           Threads & 32 & 16 & 256 \\
           L1D Cache &  48K$\times$16 & 64K$\times$16 & 1K$\times$256   \\  
           L1I Cache & 32K$\times$16 & 64K$\times$16 & 1K$\times$16  \\
           L2  Cache  & 1M$\times$16 & 0.5M$\times$16 & 1.5M$\times$1 \\
           L3  Cache  & 37.5M$\times$1 & 16M$\times$1 & 3M$\times$1 \\
           Frequency & 2.5GHz & 2.133GHz & 1.8GHz \\
        \bottomrule   
        \end{tabular}  
    }
    \label{tab_compute_resources}
    \vspace{-3ex}
\end{table}

\noindent{\bf Benchmarking Methodology: }
\squishlist
    \item First, we benchmark the general-purpose computing power in a BF3-attached server regarding memory subsystem and computing ability (\S~\ref{section_cpu}). 
    \item Second, we benchmark the networking ability of the three processors (\S~\ref{section_networking}). 
    \item Third, we present three case studies to demonstrate our hints related to how to fully exploit the potentials of DPA-enhanced SmartNIC (\S~\ref{sec_case_study}). 
\squishend

\vspace{-1ex}
\section{Benchmarking General-purpose Computing Power}
\label{section_cpu}
In this section, we characterize the general-purpose computing power in a BF3-attached server, i.e., the three processors. Prior works~\cite{battle_bf3_hoti,bf2_compression_bench,dpubench,fudan_bf2,ipipe,bf2_performance} evaluate the SmartNIC processors from a high-level application perspective. To thoroughly understand the characteristics of these processors, we assess them from two architectural perspectives: 1) computing and 2) memory subsystem.

\subsection{Benchmarking Three Computing processors}
\begin{figure}[]
    \subfloat[Host Roofline Model]{
        \label{fig_e_roofline_host}
        \includegraphics[keepaspectratio=true, width=0.49\linewidth]{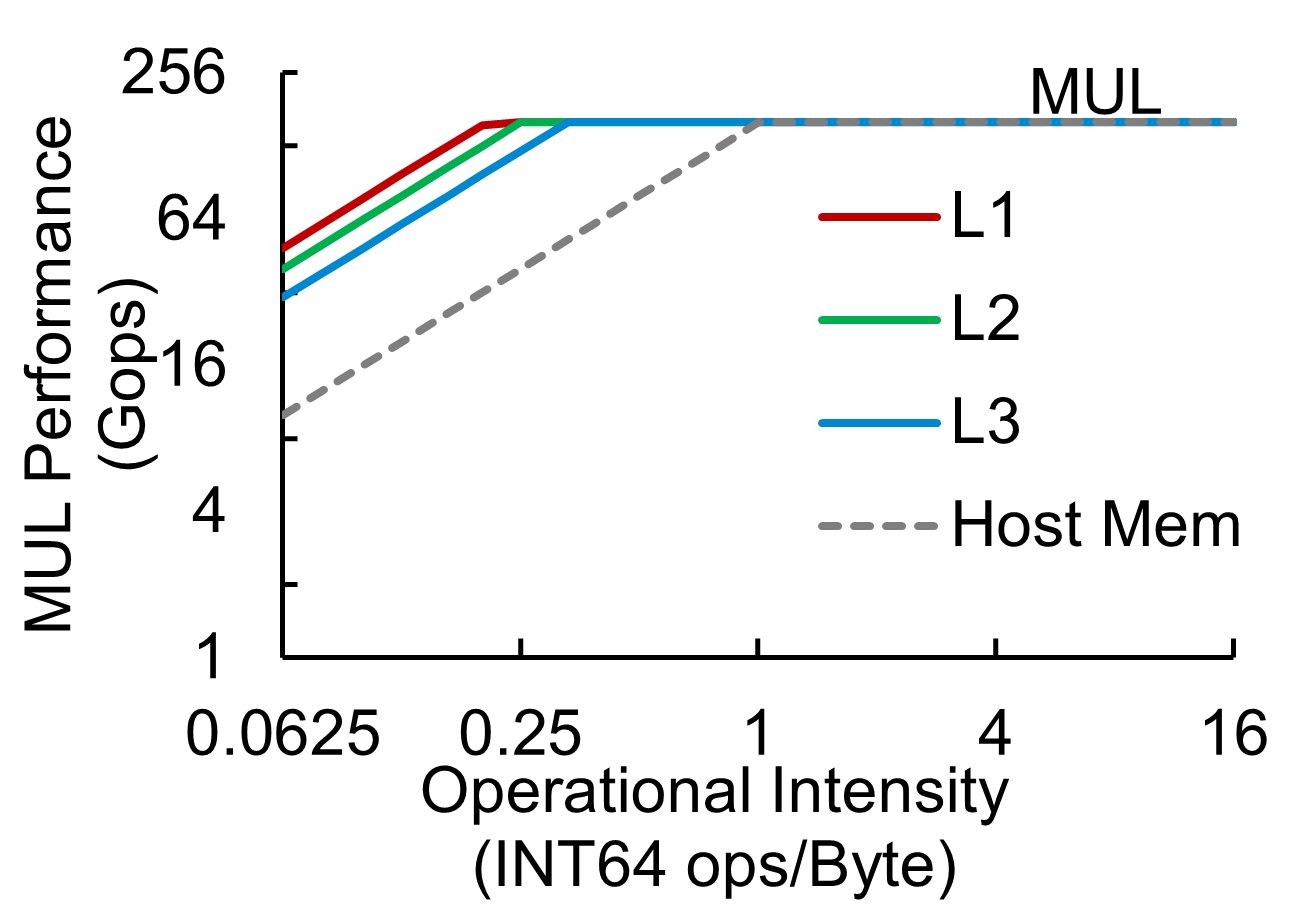}
    }
    \hspace{-3ex}
    \subfloat[Arm Roofline Model]{
        \label{fig_e_roofline_arm}
        \includegraphics[keepaspectratio=true, width=0.49\linewidth]{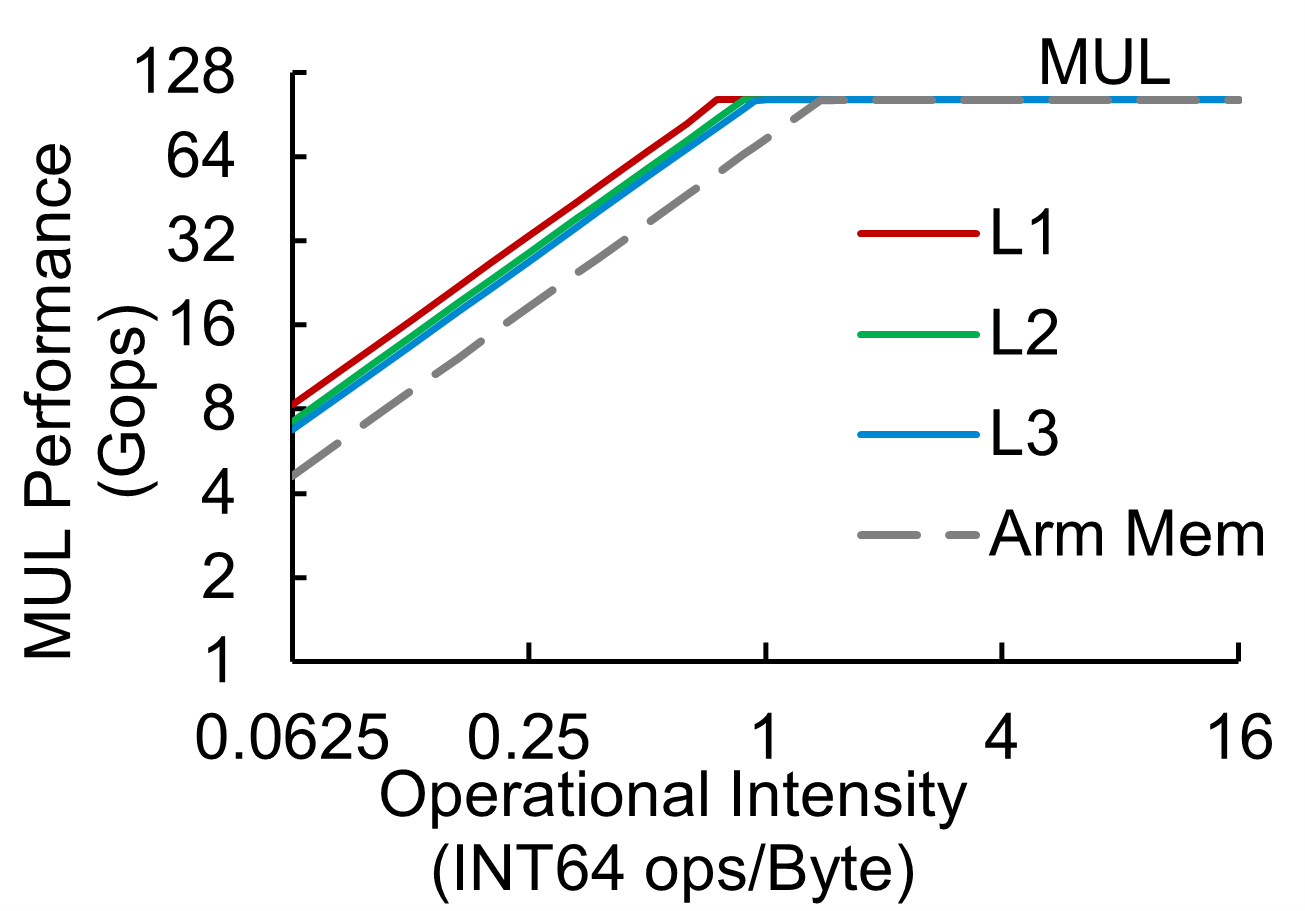}
    }
    \vspace{-3ex}
    \newline
    \subfloat[DPA Roofline Model]{
        \label{fig_e_roofline_dpa}
        \includegraphics[keepaspectratio=true, width=0.49\linewidth]{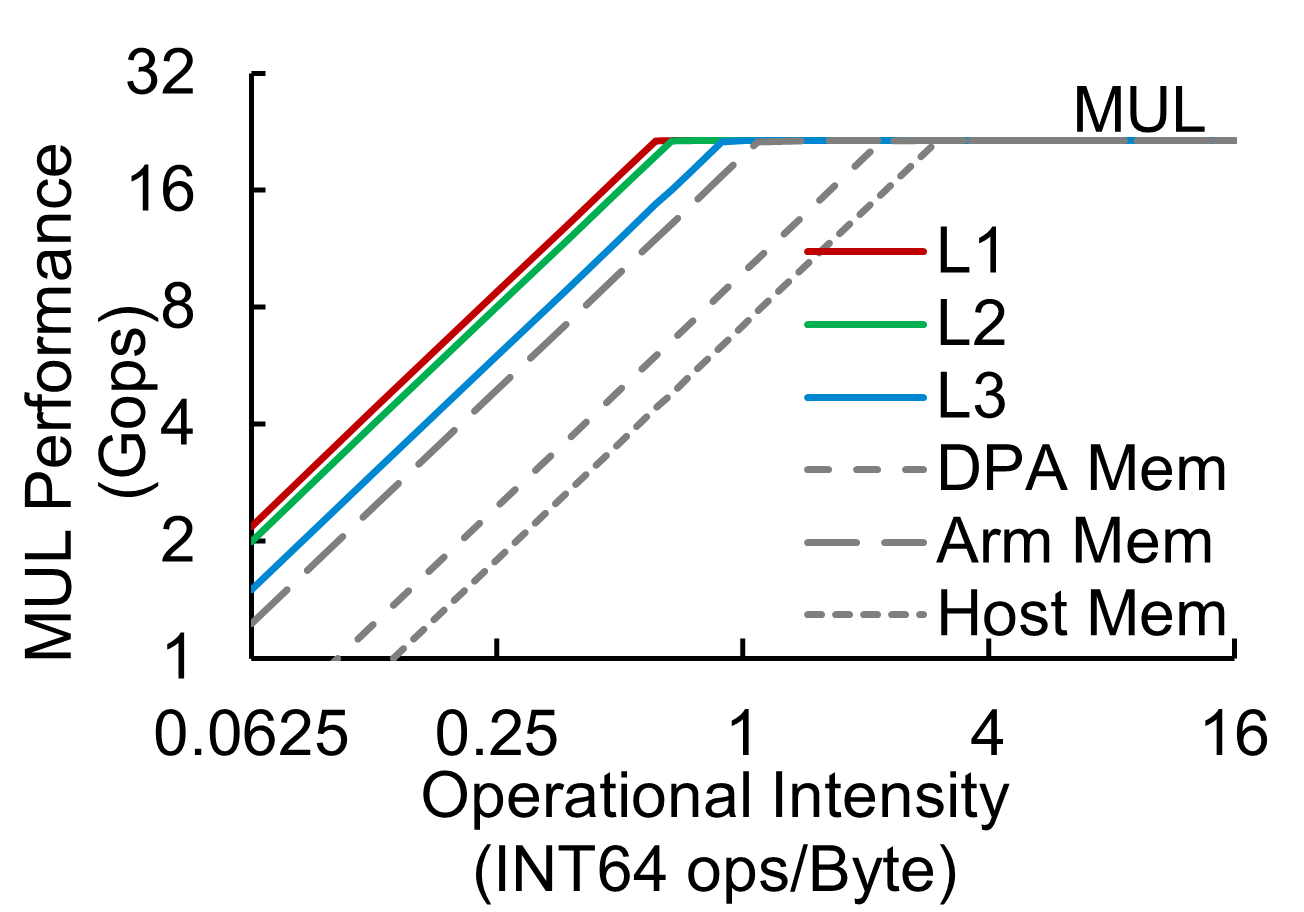}
    }
    \hspace{-3ex}
    \subfloat[Peak MUL performance for different number of DPA threads]{
        \label{fig_e_roofline_dpa_threads}
        \includegraphics[keepaspectratio=true, width=0.49\linewidth]{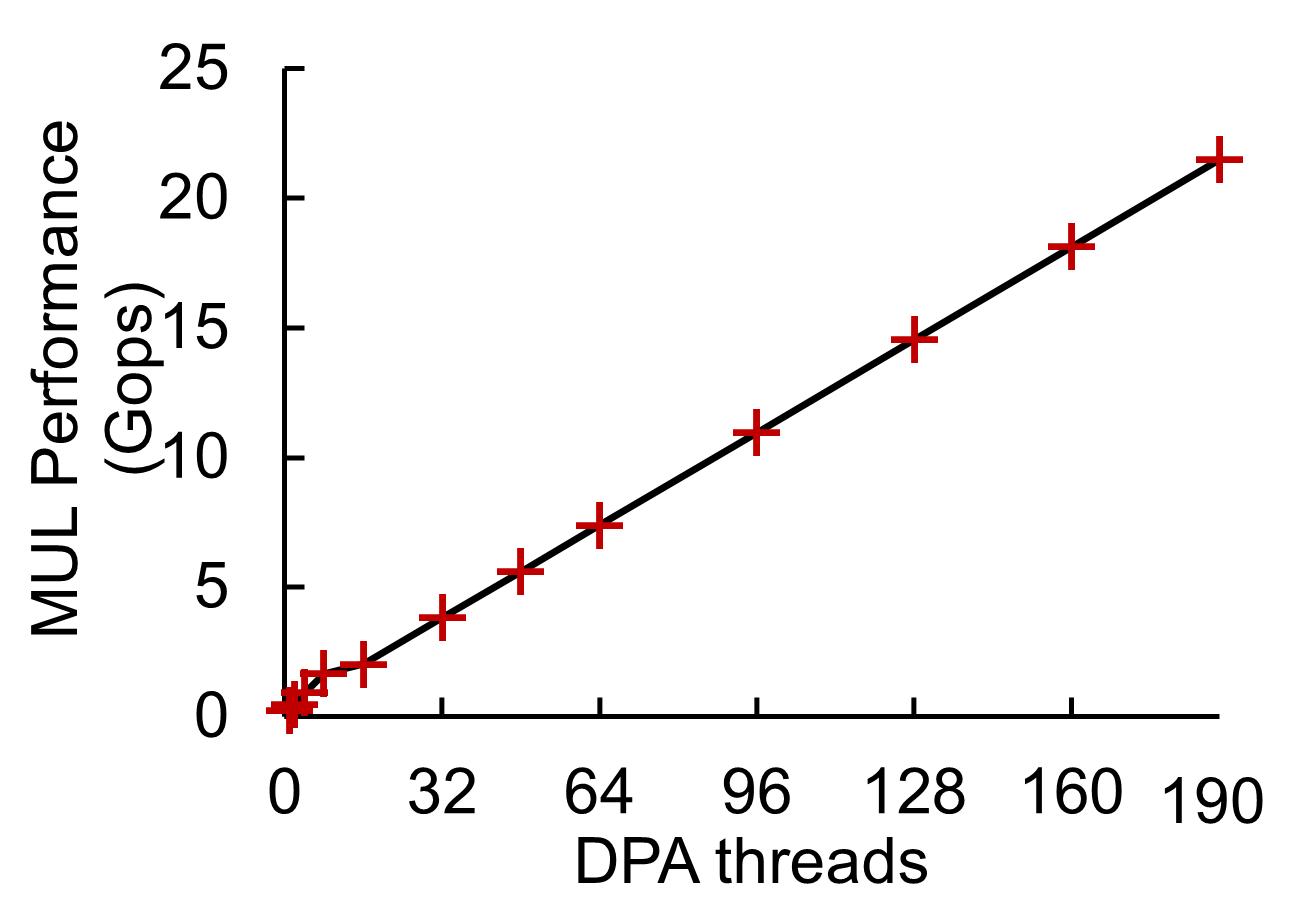}
    }
    
    \caption{Cache-aware Roofline Model for different general-purpose computing power.} 
    \label{fig_e_roofline_host_arm_dpa}
    \vspace{-3ex}
\end{figure}

In this section, we benchmark the three processors regarding their computing ability. \sideword{R3}\zheng{To experimentally assess their efficiency, we conducted a series of tests following the methodology outlined in~\cite{cache_aware_roofline}. The bandwidth is analyzed by varying the number of memory operations to access different levels of the memory hierarchy, specifically by using contiguous memory and incrementally increasing the working set sizes to fit the capacities of L1, L2, L3 caches, and main memory. Figure~\ref{fig_e_roofline_host}/~\ref{fig_e_roofline_arm}/~\ref{fig_e_roofline_dpa} shows three processors' Cache-aware Roofline Model~\cite{cache_aware_roofline} for the ``INT64 Multiplication'' operations.} We find that Arm can provide similar operations per second (Gops) comparable to that of the host CPU under the same core counts (16) and without hyper-threading. Although DPA has 256 threads (16 cores), its achievable Gops is 7.5$\times$ lower than the host CPU and 4.7$\times$ lower than the Arm. DPA's single-thread computing power is much lower than the host/Arm (up to 26$\times$ lower). \sideword{R3}\zheng{Figure~\ref{fig_e_roofline_dpa_threads} uses the same testing method as previous experiments.} It shows DPA's ``INT64 Multiplication" performance under different numbers of DPA threads, demonstrating linear scalability.

\noindent{\bf Takeaways (Computing):} Arm cores can serve as a powerful supplement to the host's computing capabilities. DPA's single-thread computing power is very low, and serial compute-intensive workloads should not be offloaded to DPA cores.

\noindent{\bf DPA's Unique Computing Characteristic:} Unlike the host/Arm, DPA has a thread count (256) that is an order of magnitude higher. To fully leverage DPA's underperforming computing ability, the workloads should be very easy to parallelize. We further use a case study to explore this DPA's unique computing characteristics in~\S~\ref{section_case_study_nf}.

\subsection{Memory Subsystem}
In this section, we evaluate the memory subsystem of the BF3-attached server. All three processors in the BF3-attached server have their own three-level cache. Both X86 and Arm core have directly connected DDR memory.

\begin{figure}
	\centering
	{\includegraphics[keepaspectratio=true, width=0.9\linewidth]{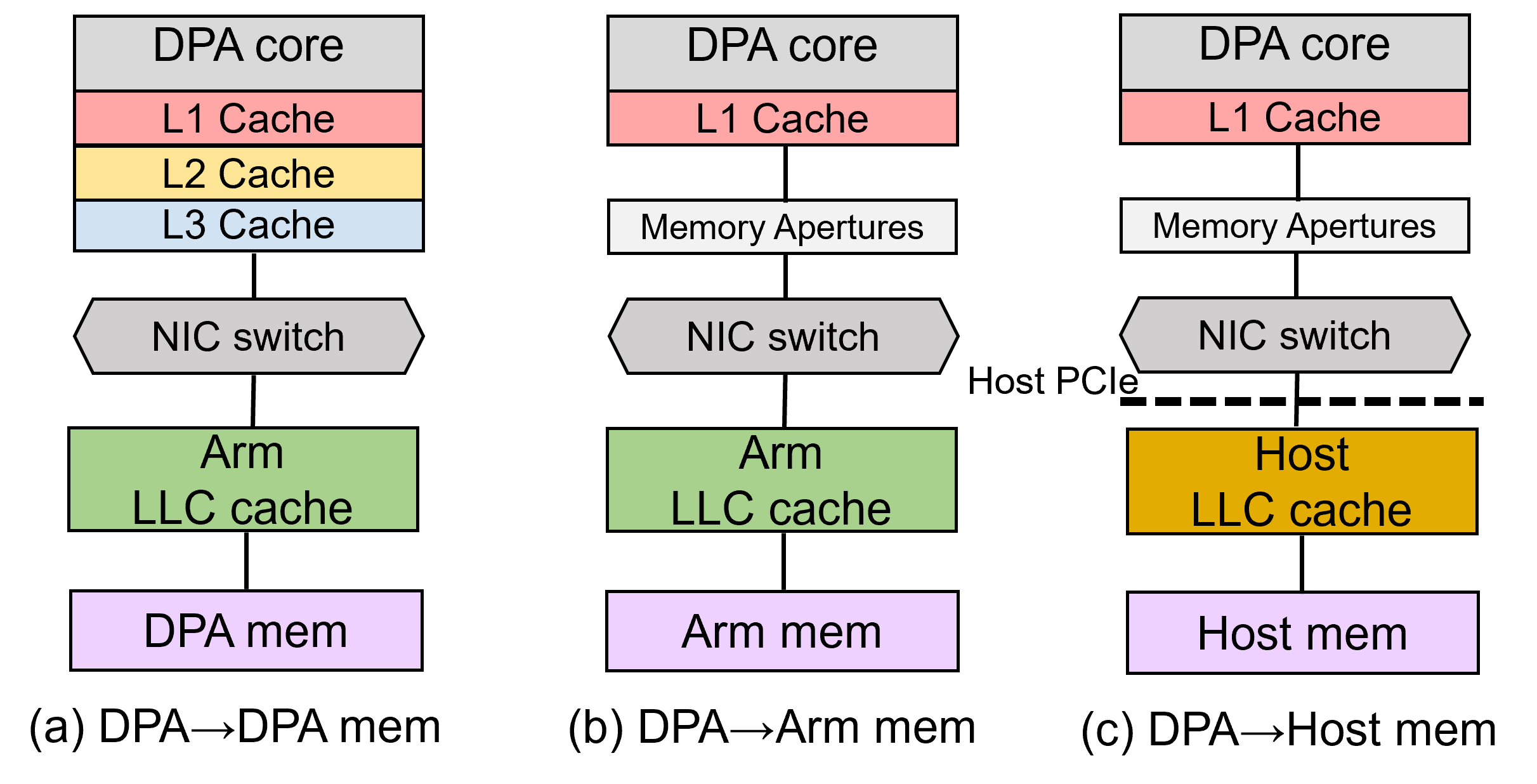}

	\caption{DPA accesses three memory types.}
	\label{fig_dpa_cache_policy}}
    \vspace{-3ex}
\end{figure}

DPA does not have directly connected DDR memory, instead, a 1-GB region in the Arm DDR is allocated exclusively for DPA access. DPA has to access this memory region (DPA memory) through the NIC switch which incurs high latency. DPA's access to the DPA memory would be cached with Arm's L3 cache and DPA's L1/L2/L3 cache. In addition to DPA memory, DPA can access the host memory and Arm memory through a memory aperture module~\cite{dpa_programming_guide} near the DPA core. This module converts a memory request into a PCIe transaction, thus allowing DPA cores to directly access Arm memory and host memory using a load/store instruction. It's worth noting that such accesses will only go through DPA's L1 cache and will not go through DPA's L2/L3 caches. 
Also, DPA's access to the host/Arm memory would go through the host L3 cache or Arm L3 cache. Figure~\ref{fig_dpa_cache_policy} shows the physical paths of the DPA's access to these three kinds of memories. 

\begin{figure}
	\centering
	{\includegraphics[keepaspectratio=true, width=0.95\linewidth]{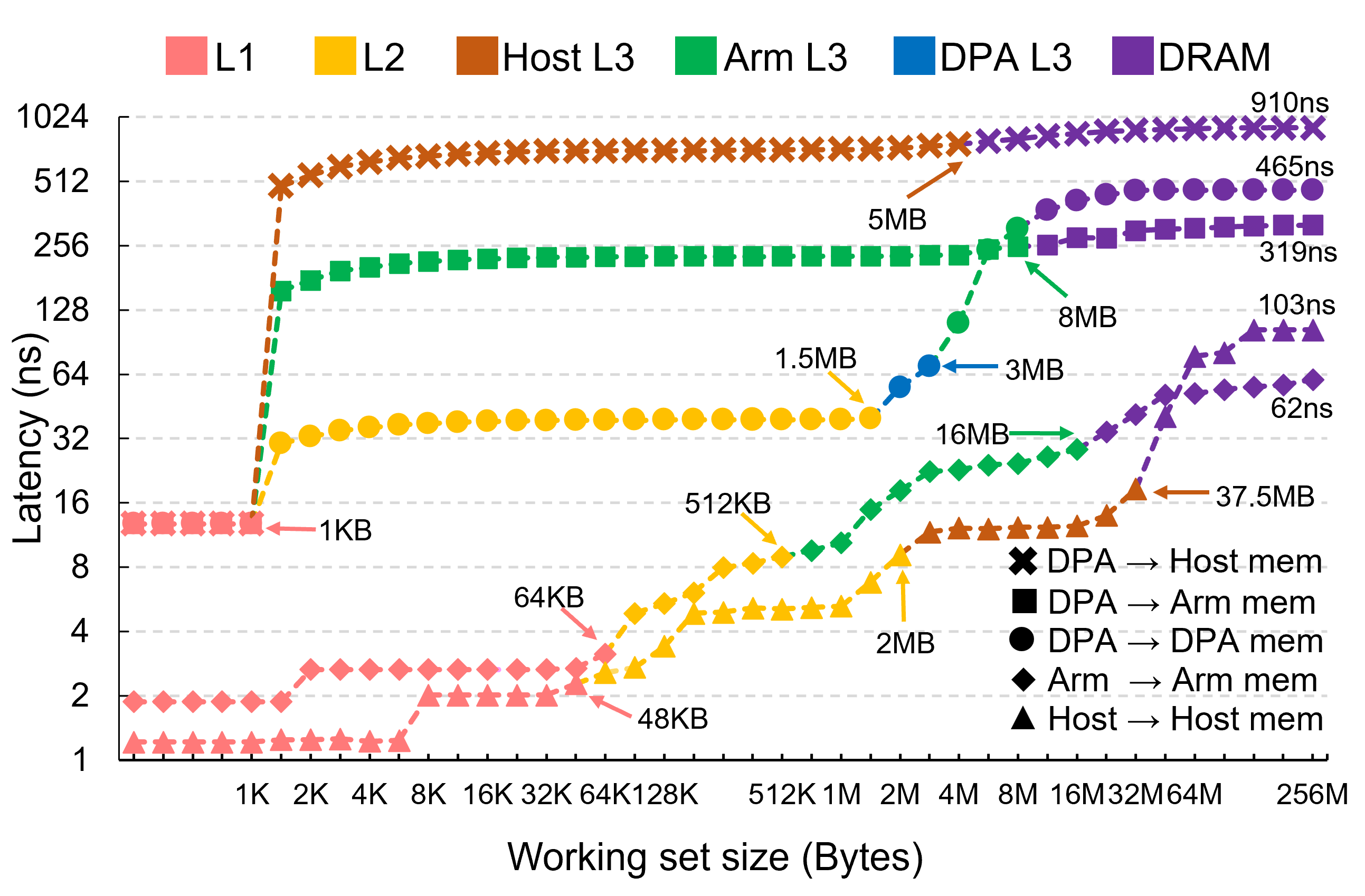}
    \vspace{-1ex}

    \caption{Cache latency for all computer resources.}
    \label{fig_e_cache_latency}}
    \vspace{-2.5ex}
\end{figure}

\subsubsection{Cache/Memory Latency} 
We first measure the memory/cache read latency. We use a pointer-chasing manner~\cite{k_pointer_chasing} and carefully vary the access stride and the working set size. Figure~\ref{fig_e_cache_latency} shows the latencies of the supported five kinds of memory accesses. ``X$\rightarrow$Y mem'' means X processor accesses Y memory. We have three observations.

First, The L1 cache latency of DPA is 10.5$\times$ of the host L1 latency. DPA's L2/L3 cache latency is also significantly higher than that of the host/Arm. The high cache latency of DPA indicates that carefully using DPA cores is extremely important for application offloading.

Second, DPA has noticeably higher memory reading latency than Arm and the host. DPA's memory read latency is at least several times higher than the host/Arm. This can primarily be attributed to the high latency of the NIC switch and PCIe interconnect. Compared to the Arm core and host core's direct memory access, DPA's access to the DPA memory or Arm memory would involve an additional NIC switch. In addition to the NIC switch, DPA's access to host memory also involves another PCIe interconnect.

Third, although Arm memory and DPA memory are physically the same DDRs connected to the Arm core, the latency of DPA's access to the DPA memory is noticeably higher than DPA's access to the Arm memory. We suspect this is mainly because DPA's access to the host memory and Arm memory only go through DPA's L1 cache~\cite{dpa_programming_guide}, while DPA's access to the DPA memory would additionally go through DPA's poor L2/L3 cache. The latency of ``DPA$\rightarrow$Host'' is still higher than DPA accessing DPA memory although it bypasses the L2/L3 cache, due to the additional step of the host PCIe interconnect.

\subsubsection{Cache Bandwidth}
In this section, we measure DPA's random read bandwidth from a DPA memory buffer. We vary the memory buffer size (i.e., the working set size) and use different numbers of threads.

\begin{figure}[]
    \subfloat[Per-thread]{
        \label{fig_e_memory_dpa_mem_random_access_1_thread}
        \includegraphics[keepaspectratio=true, width=0.49\linewidth]{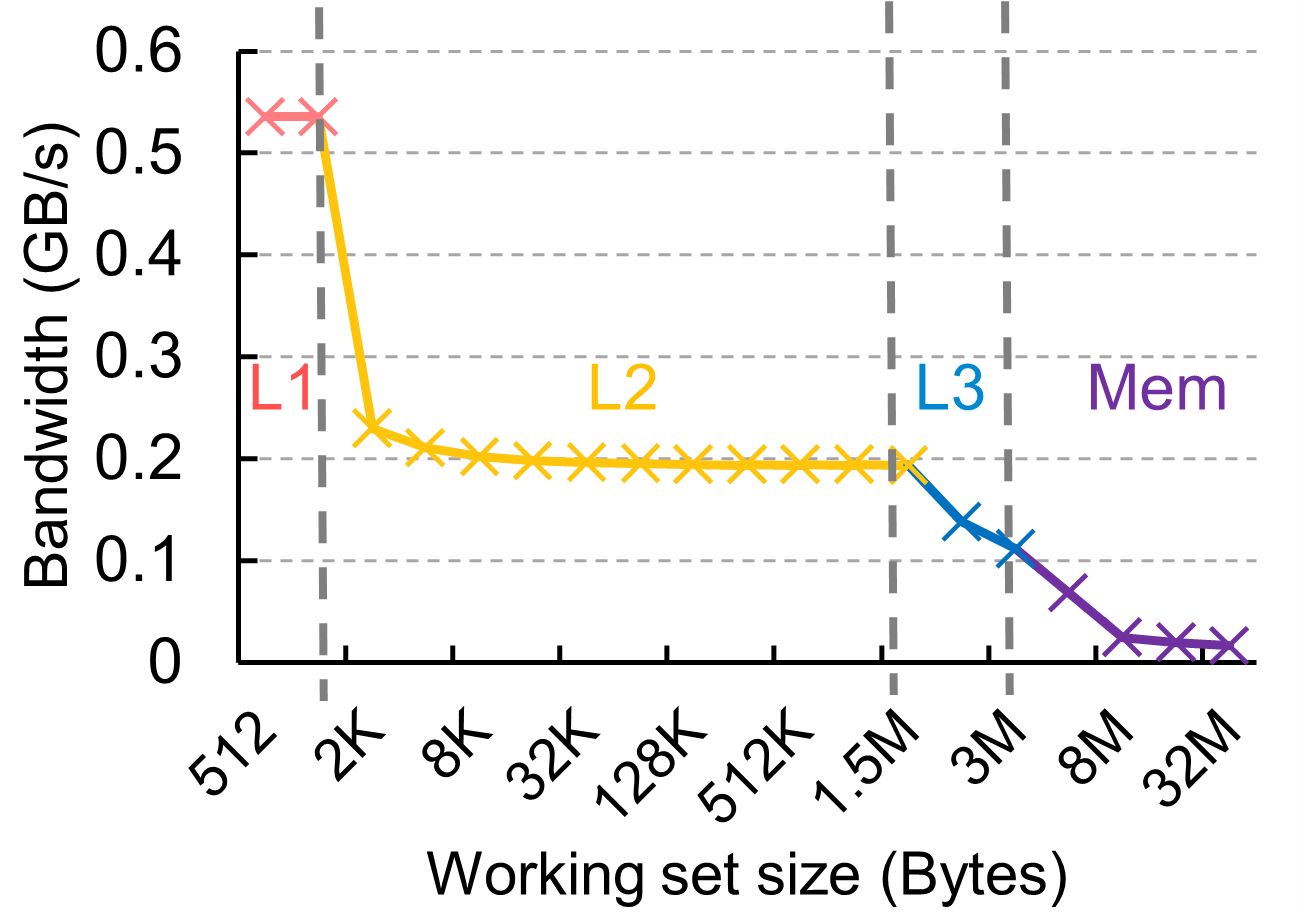}
    }
    \subfloat[All-threads]{
        \label{fig_e_memory_dpa_mem_random_access_190_threads}
        \includegraphics[keepaspectratio=true, width=0.49\linewidth]{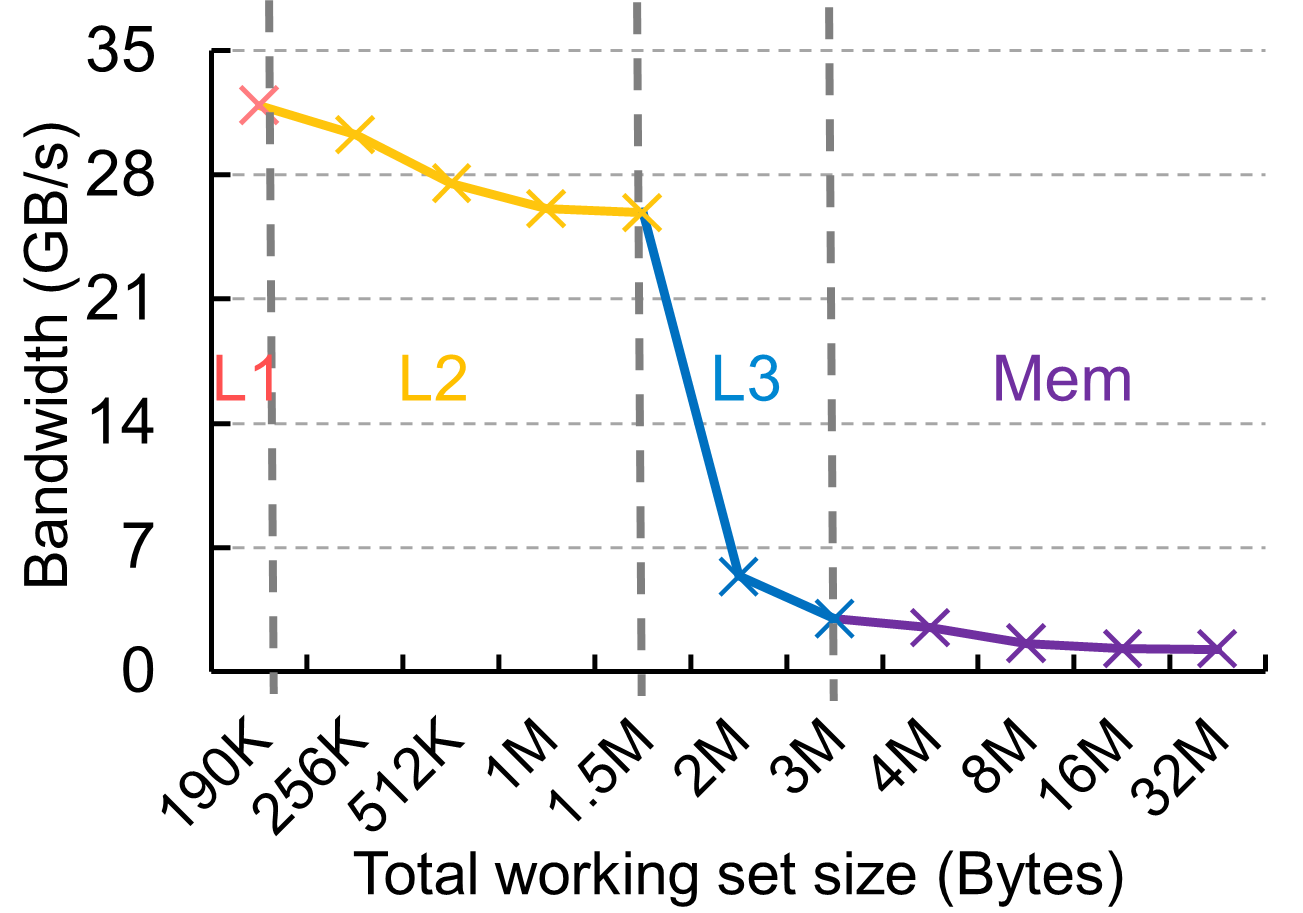}
    }
    \caption{DPA random access DPA memory throughput.}
    \label{fig_e_memory_dpa_mem_random_access}
    \vspace{-3ex}
\end{figure}

Figure~\ref{fig_e_memory_dpa_mem_random_access_1_thread} shows the read bandwidth of using a single DPA thread and Figure~\ref{fig_e_memory_dpa_mem_random_access_190_threads} shows the bandwidth of using all 190 DPA threads. We observe that the read bandwidth significantly drops when the working set exceeds DPA's L2 cache size (1.5 MB), the bandwidth loss can be up to 25$\times$. This indicates that memory-intensive DPA applications must carefully manage their working set size.

\subsubsection{Memory Bandwidth}
\label{section_memory_bandwidth}
\begin{figure}[]
    \subfloat[Per-thread]{
        \label{fig_e_memory_bandwidth_per_thread}
        \includegraphics[keepaspectratio=true, width=0.49\linewidth]{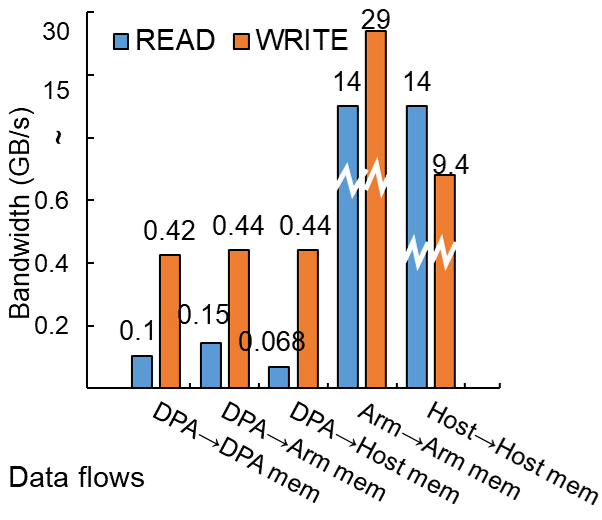}
    }
    \subfloat[All-threads]{
        \label{fig_e_memory_bandwidth_all_thread}
        \includegraphics[keepaspectratio=true, width=0.49\linewidth]{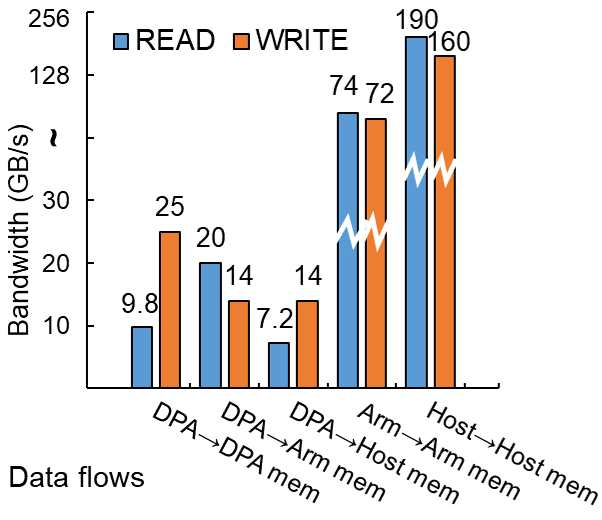}
    }
    \caption{Achievable memory throughput.} 
    \label{fig_e_memory_bandwidth}
    \vspace{-3ex}
\end{figure}

Figure~\ref{fig_e_memory_bandwidth} shows the per-thread and all-thread sequential read bandwidth of the mentioned five kinds of accesses. We let the working set be large enough that all reading requests are served by the memory instead of the cache. ``X$\rightarrow$Y mem'' means X cores access Y memory. The all-threads bandwidth is measured by using all threads concurrently. We have two observations. 

First, no matter access which memory, DPA has up to 205$\times$ lower per-thread read/write bandwidth than the host and Arm. Although DPA has 256 threads, the all-threads bandwidth is still up to 7.6$\times$ lower than Arm and the host. The all-thread bandwidth is also much lower than the network line rate (400 Gbps full-duplex). This implies that DPA can perform poorly in memory-intensive applications.

    Second, Arm cores in BF3 have a comparable per-thread memory bandwidth to host cores. This suggests offloading memory-intensive workloads to BF3's Arm cores. The all-threads memory bandwidth of the host is 2.7$\times$ higher than the Arm because BF3's Arm only has two memory channels while the host features eight memory channels.

\begin{figure}[]
    \subfloat[Mixed read bandwidth of \\``DPA mem + Arm mem'']{
        \label{fig_e_memory_concurrency_dpa_arm_read}
        \includegraphics[keepaspectratio=true, width=0.47\linewidth]{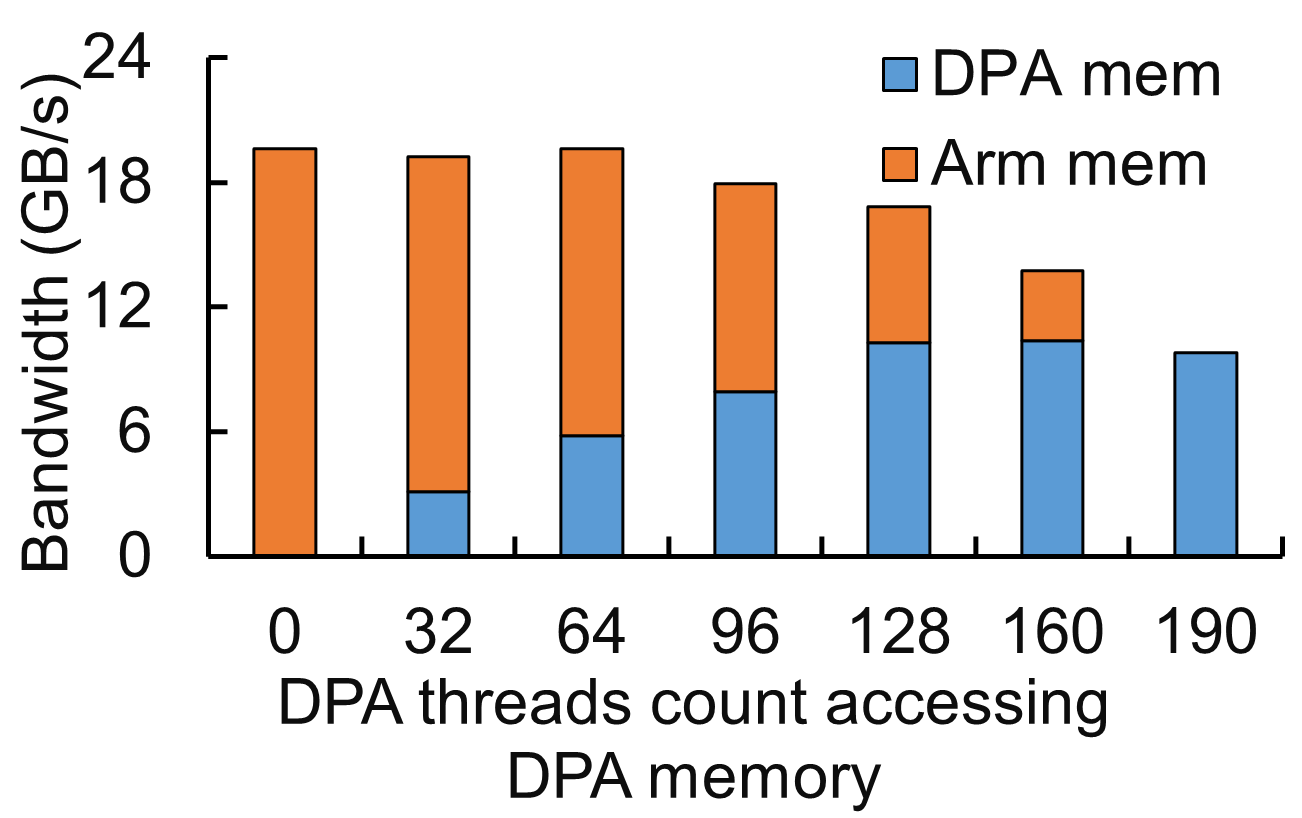}
    }
    \subfloat[Mixed write bandwidth of \\``DPA mem + Arm mem'']{
        \label{fig_e_memory_concurrency_dpa_arm_write}
        \includegraphics[keepaspectratio=true, width=0.47\linewidth]{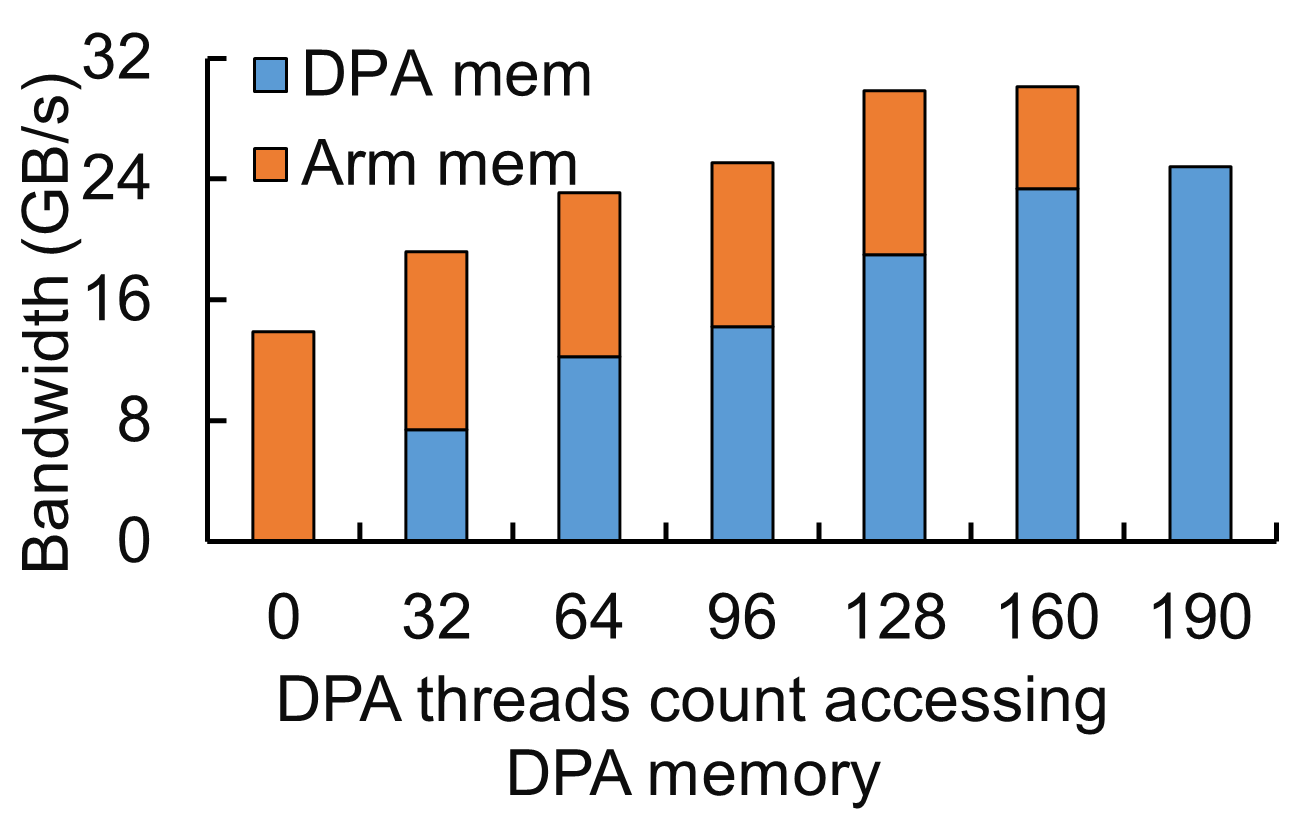}
    }
    \vspace{-1ex}
    \newline
    \subfloat[Mixed read bandwidth of \\``DPA mem + Host mem'']{
        \label{fig_e_memory_concurrency_dpa_host_read}
        \includegraphics[keepaspectratio=true, width=0.47\linewidth]{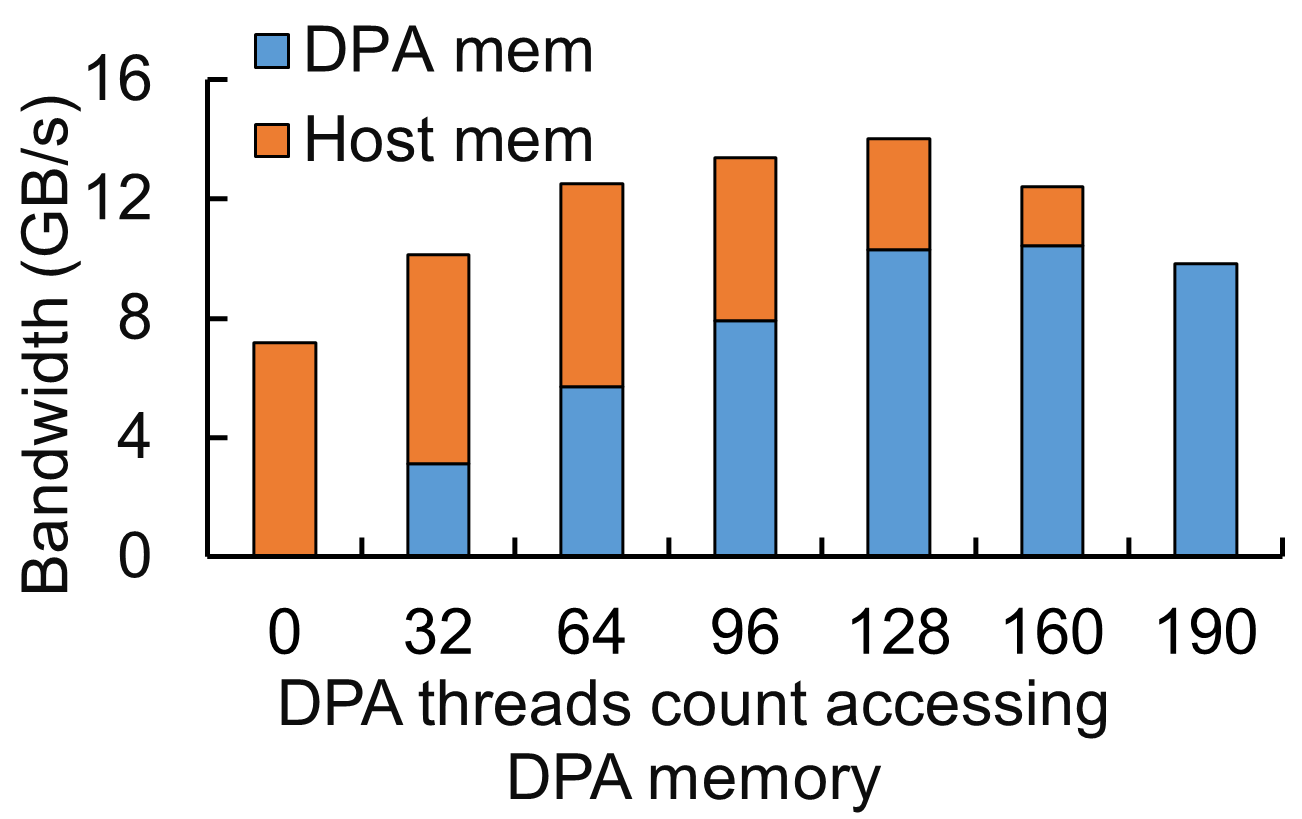}
    }
    \subfloat[Mixed write bandwidth of \\``DPA mem + Host mem'']{
        \label{fig_e_memory_concurrency_dpa_host_write}
        \includegraphics[keepaspectratio=true, width=0.47\linewidth]{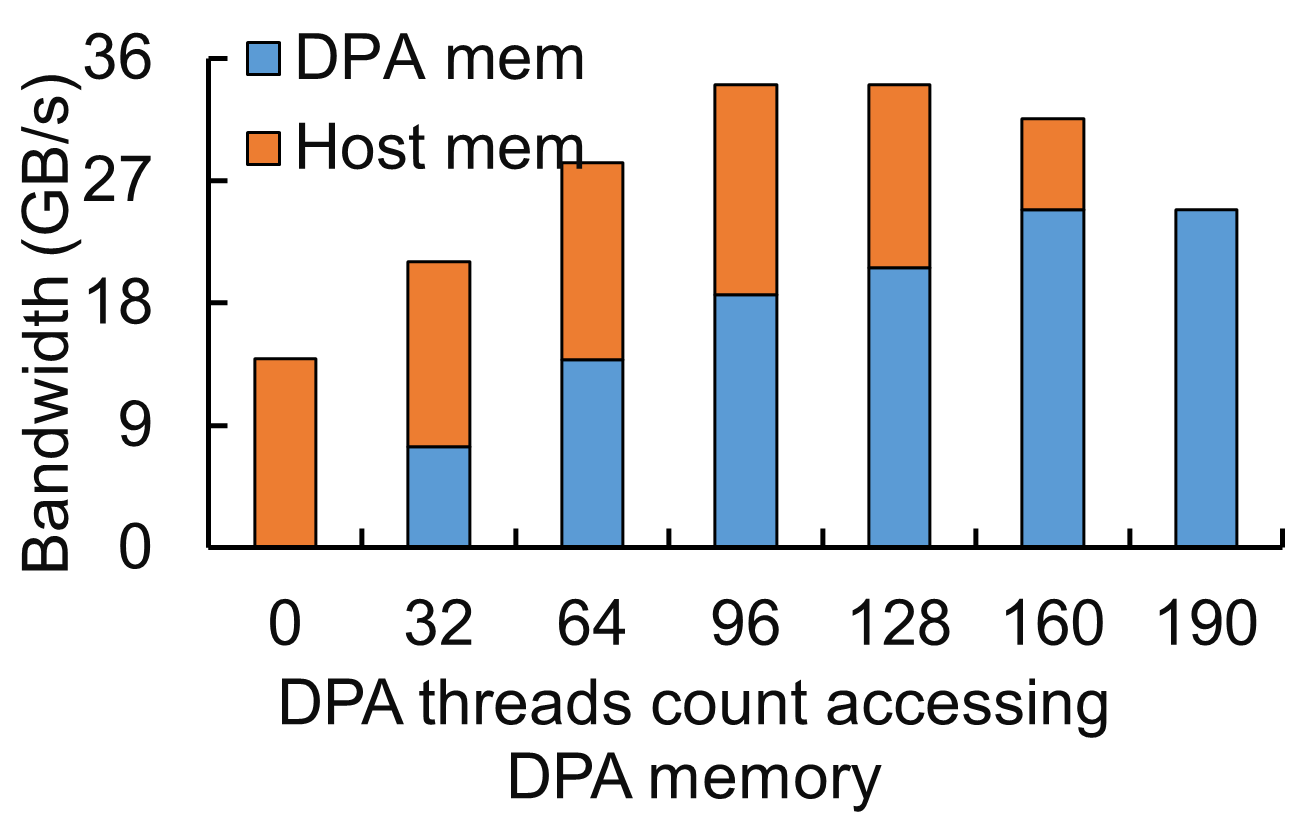}
    }    
    \caption{Mixed memory throughput under 190 available DPA threads. We let different numbers of DPA threads access DPA memory and let all remaining DPA threads access the Host/Arm memory.} 
    \label{fig_e_memory_concurrency}
    \vspace{-3ex}
\end{figure}

\noindent{\bf Mixed memory bandwidth of DPA cores.}
From Figure~\ref{fig_e_memory_bandwidth}'s DPA-related results, we can observe that the all-threads memory bandwidth for each memory is far lower than the product of per-thread bandwidth and the number of DPA threads. This indicates that the DPA's memory bandwidth is not bounded by the DPA thread count. This inspires us to explore whether we could use the spare DPA threads to access other different memories to further improve DPA's all-threads memory bandwidth.

Accordingly, we use a different number of DPA threads to access DPA memory and the remaining threads to access the host memory or Arm memory\footnote{We do not let DPA access the host memory and Arm memory concurrently because the current DOCA framework does not allow this kind operation.}. Figure~\ref{fig_e_memory_concurrency} shows the mixed read/write bandwidth for two combinations: ``DPA mem + Arm mem'' and ``DPA mem + Host mem''. We observe that the write bandwidth of ``DPA mem + Arm mem'' and the read/write bandwidth of ``DPA mem + Host mem'' are all higher than only using a single type of memory. The maximum bandwidth improvement can be up to 2.4$\times$. This indicates that memory-intensive workloads running in DPA cores should consider using multiple kinds of memories concurrently to maximize the achievable memory bandwidth.

\noindent{\bf Takeaways (Memory subsystem).} DPA's cache/memory latency is much higher than the host/Arm. DPA's all-threads memory bandwidth is also significantly lower than the host/Arm. 
Unlike the host/Arm, DPA does not have a directly connected memory. DPA has to access DPA memory or Arm memory through the NIC switch, which can greatly limit the performance on memory-intensive workloads. As such, DPA applications' working set sizes should better fit in the DPA cache size. DPA can use multiple kinds of memories concurrently to improve achievable memory bandwidth. We further use a case study to explore the influence of DPA's unique memory characteristic in~\S~\ref{section_case_study_kv_aggr}.

\vspace{-1ex}
\section{Benchmakring Networking}
In this section, we evaluate the networking ability of a BF3-attached server. 
\label{section_networking}
The datapath accelerator enhanced BF3 offers various strategies to leverage its powerful networking ability. The host, DPA, and Arm cores can send/receive network packets. In this section, we use two back-to-back connected servers as mentioned in~\S~\ref{sec_server_setup} to evaluate the networking abilities of the BF3-attached server.

\subsection{Effect of Network Buffer Location}
We first examine the effects of network buffer location (either host/Arm memory or DPA memory) when using DPA core to process network traffic\footnote{We do not explore the host CPU/Arm's mechanism since there are plenty of direct cache access related researches~\cite{k_ddio_1,k_ddio_2,k_ddio_3} about X86/Arm.}. 

\noindent{\bf Host/Arm Memory.}
When DPA cores use the host/Arm memory as the network buffer, NIC can directly put the newly arrived packets into the host/Arm L3 cache. Regarding processing latency, letting DPA cores use the host/Arm memory as the network buffer is not a good choice, because NIC writing packets into the host/Arm L3 cache would travel through the high-latency NIC switch (and an additional PCIe interconnect if using host memory). We quantitatively analyze this impact on the network latency in~\S~\ref{section_network_latency}. 

\noindent{\bf DPA Memory.}
When DPA cores use DPA memory as the network buffer, NIC can directly put the newly arrived packets into the DPA L2 or L3 cache. 

To validate the effect of network buffer location, we conduct the following experiment. 
We first sequentially send a bunch of packets (e.g., 512 packets of 1KB each) to a DPA core that uses DPA memory as the network receives buffer. After a long enough period to make sure all packets have arrived, the DPA core reads the first packet of this bunch of packets and measures the read latency. Then we sequentially send another bunch of packets, and the DPA core measures the latency of the second packet. \sideword{R3}\zheng{We repeat this process until we get the latency of the last packet. Note that we only measure one packet latency to prevent the effects of cache prefetching, in addition, we would pollute the DPA cache before the sending side sends a bunch of packets.}

\begin{figure}
	\centering
	{\includegraphics[keepaspectratio=true, width=0.9\linewidth]{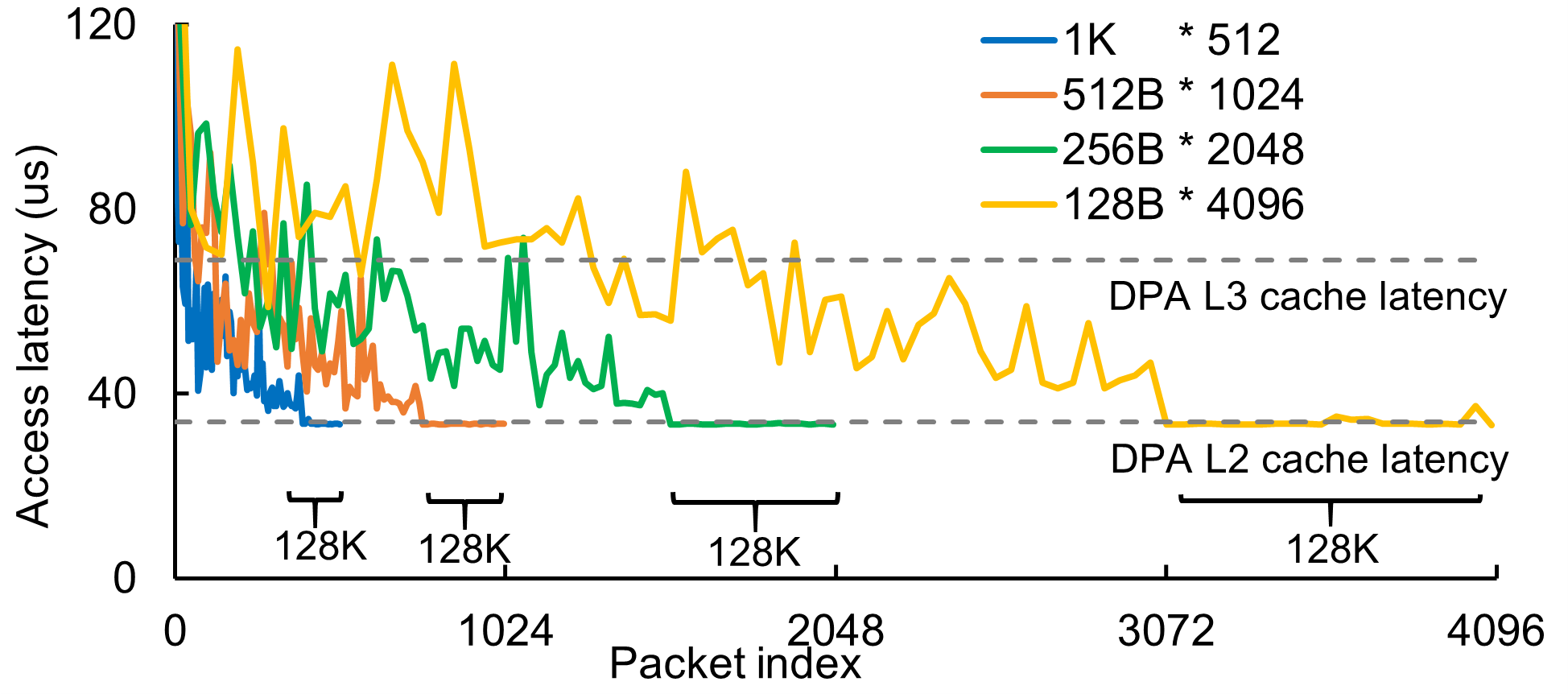}
    \vspace{-1ex}

	\caption{ Average packet access latency under different combinations with a total of 512KB size. A higher index corresponds to a more recently received packet.}
	\label{fig_e_network_dpa_ddio_working_set}}
    \vspace{-3ex}
\end{figure}

Figure~\ref{fig_e_network_dpa_ddio_working_set} shows the access latency of each packet. We have two observations. 
First, the latest received 128 KB packets are guaranteed to be put in the DPA L2 cache. No matter the packet size, the access latency of the latest 128 KB packets is always DPA L2 latency, indicating that these packets are directly put into the L2 cache by the NIC. 

Second, the newer received packet has a higher possibility of being put into the places that are closer to the DPA core, DPA L2/DPA L3/Arm L3/DPA memory from the closest to the farthest, respectively.

\noindent{\bf Takeaways (Working set size).} When DPA is managing network traffic, the NIC can directly access the host L3 cache (host memory as network buffer), Arm L3 cache (Arm memory or DPA memory as network buffer), and DPA L2/L3 cache (DPA memory as network buffer). Therefore, programmers need to be aware of the working set size such that the working set stays at the expected cache and thus the memory traffic can be minimized. 

\subsection{Network Latency}
\label{section_network_latency}
\begin{figure}
	\centering
	{\includegraphics[keepaspectratio=true, width=0.87\linewidth]{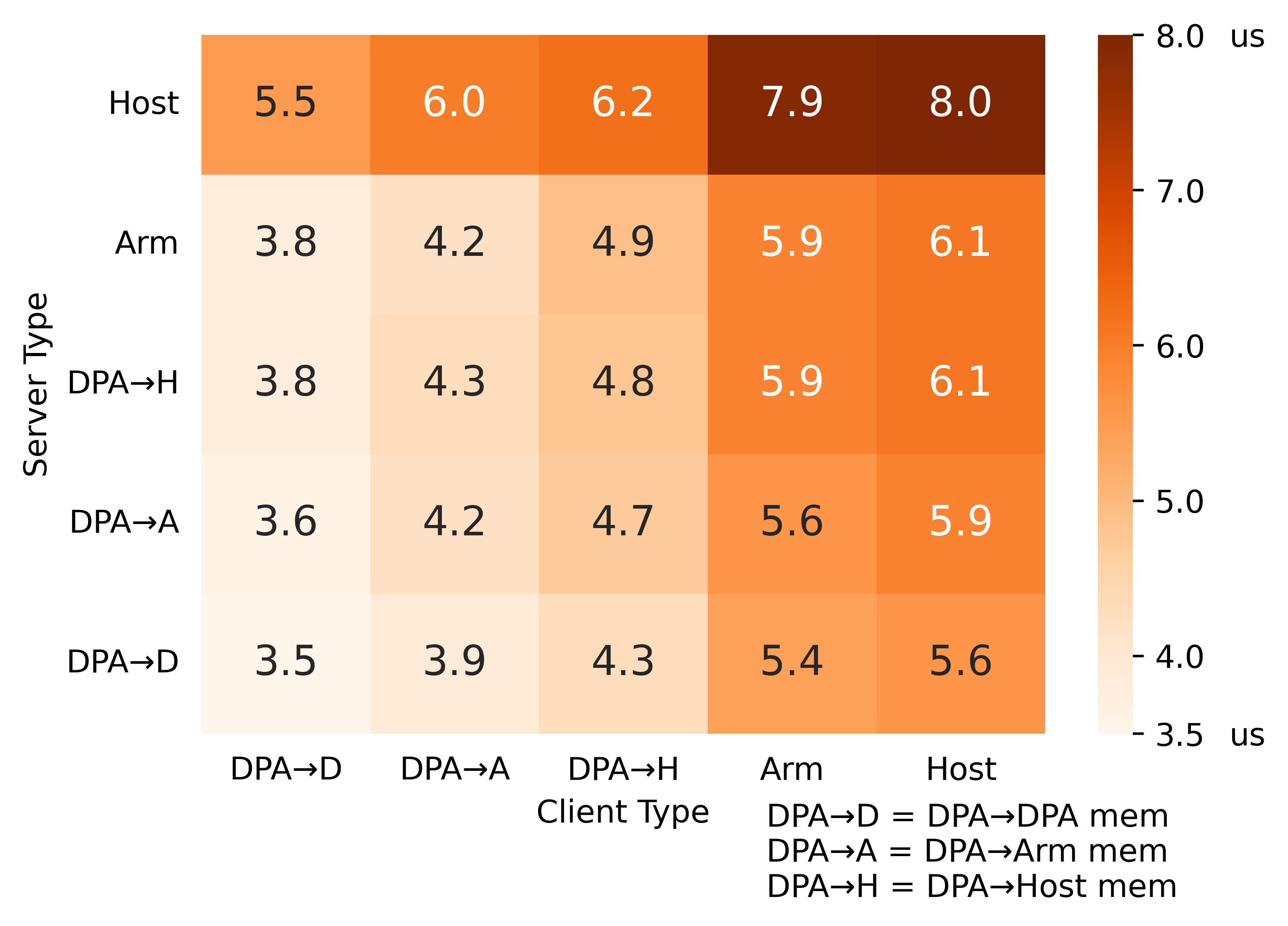}
    \vspace{-1ex}

	\caption{L2 reflector latency for different units.}
	\label{fig_e_latency_l2_reflector_5_5}}
    \vspace{-3ex}
\end{figure}

In this section, we measure the network latency regarding different computing cores via an L2-reflector application, as shown in Figure~\ref{fig_e_latency_l2_reflector_5_5}. The client sends a packet to the server. The server swaps the source and destination MAC addresses of each packet and returns the packet to the client. The packet size is fixed to 1 KB. We deploy the client/server in different processors and measure the average round-trip time, and DPA uses different types of memories. We have two observations.

First, regarding different processor implementations, the DPA-based client and server offer the lowest latency while the host-based client and server offer the highest latency. This also aligns with our expectations that DPA enjoys the shortest distance from the network because DPA and NIC are in the same NIC chip. Arm experiences moderate network latency due to its additional NIC switch hop, and the host CPU suffers from the longest network latency due to its additional NIC switch hop and its host PCIe interconnect. 

Second, the DPA implementation on DPA memory offers the lowest latency compared to that on host/Arm memory. This is because received network packets can be directly put into DPA's L2/L3 memory when using DPA memory without crossing the high-latency NIC switch or PCIe interconnect. When using the Arm memory, the received packets can only be put into the Arm L3 cache through the NIC switch. When using the host memory, the received packets can only be put into the host L3 cache through the NIC switch and PCIe interconnect.


%

\noindent{\bf Latency vs. Application complexity.}
Although DPA provides lower network latency, its memory subsystem and computing power are far inferior to the host/Arm. When a networking application involves heavy processing, the advantage of DPA being closer to the network is overshadowed by its poor memory and computing performance. 

\begin{figure}[]
    \subfloat[Access sub-packet]{
        \label{fig_net_latency_l2_reflector_access_memory_1k}
        \includegraphics[keepaspectratio=true, width=0.47\linewidth]{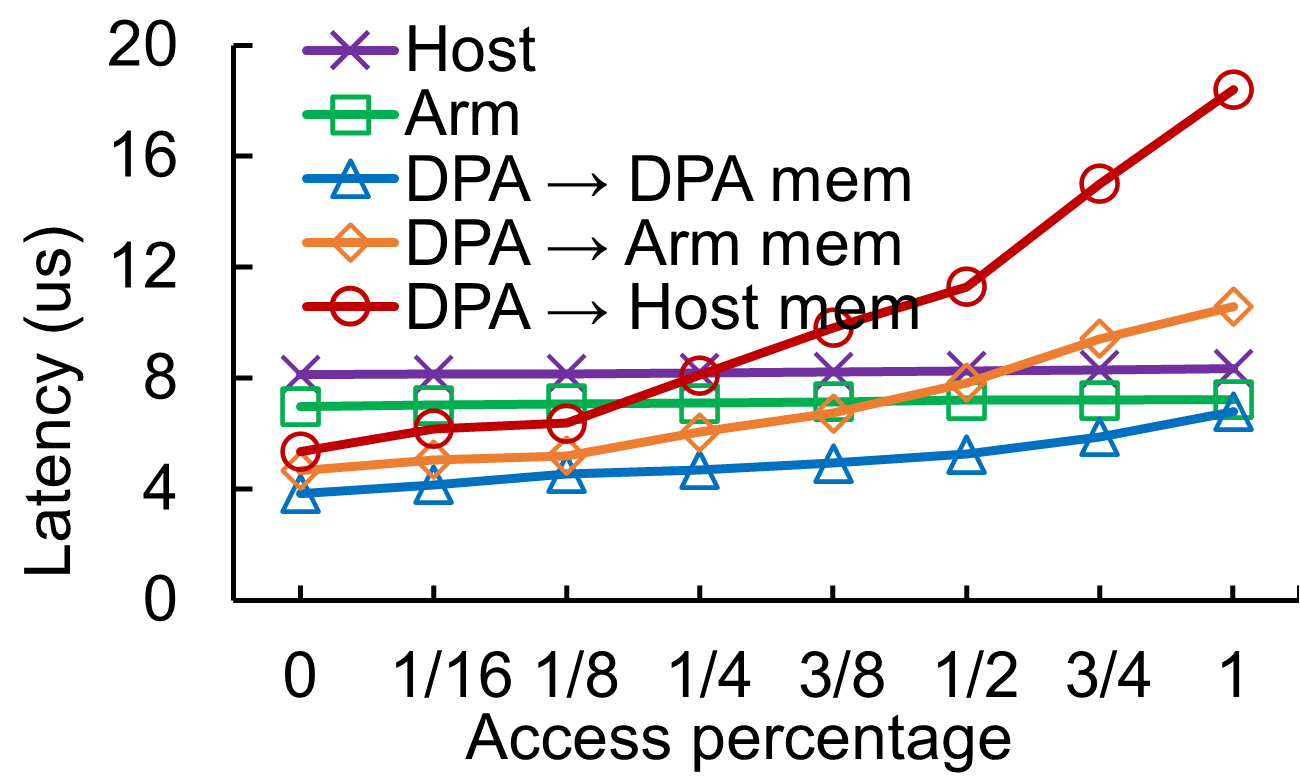}
    }
    \hspace{-1ex}
    \subfloat[Random access in 8MB]{
        \label{fig_e_net_latency_l2_reflector_access_random_8M}
        \includegraphics[keepaspectratio=true, width=0.47\linewidth]{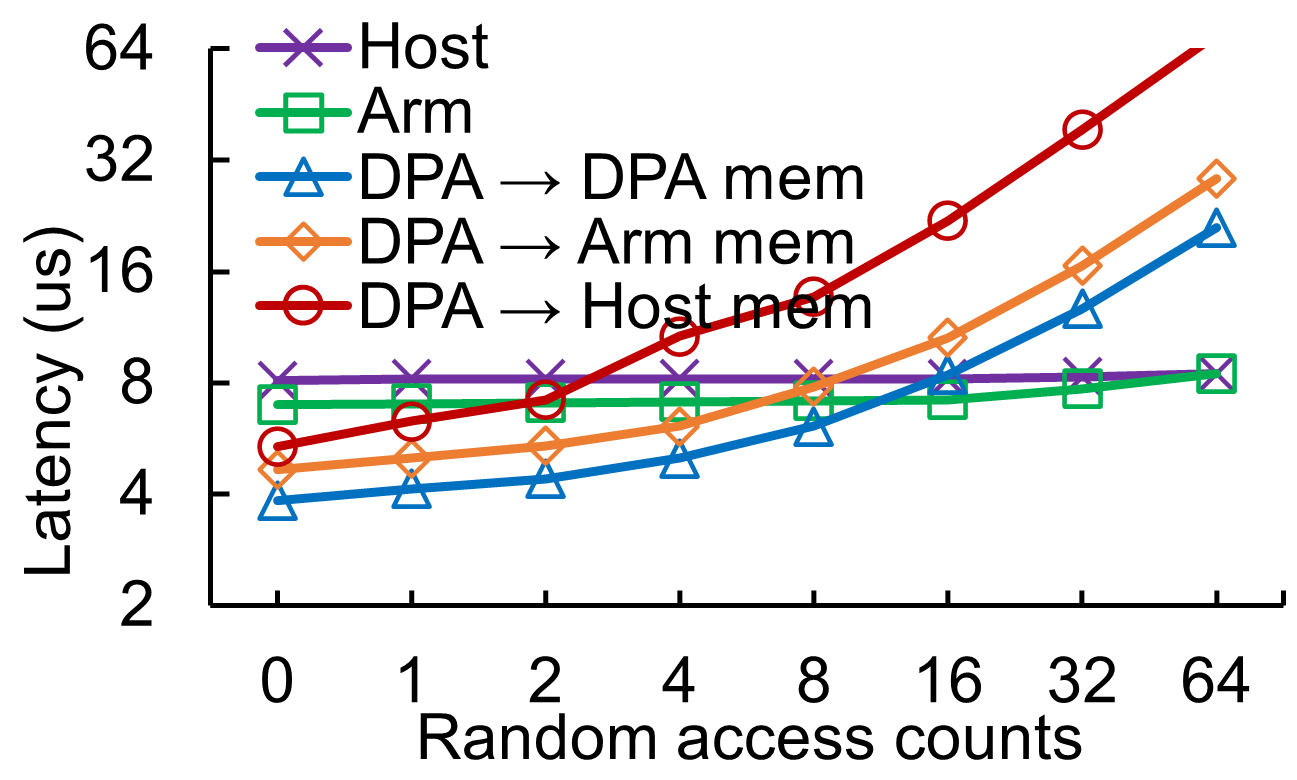}
    }
    \caption{L2 reflector latency on different memory.} 
    \label{fig_e_net_latency_l2_reflector_access_memory}
    \vspace{-3ex}
\end{figure}

We conduct two experiments to demonstrate this issue. In the first experiment, the L2 reflector server additionally reads different percentages of the packet and sums them up as a one-by-one 8-byte integer, as shown in Figure~\ref{fig_net_latency_l2_reflector_access_memory_1k}. 
In the second experiment, we approximate a memory-intensive network function by configuring the L2 reflector to perform different numbers of random memory reads per packet from an 8 MB buffer, as shown in Figure~\ref{fig_e_net_latency_l2_reflector_access_random_8M}. In both experiments, the packet size is fixed to 1 KB, and we measure the end-to-end L2-reflector latency of the host, Arm, and DPA using three different memories. 

We observe that with the increase in the reading ratio or the number of per-packet random accesses, DPA's latency significantly increases while the host/Arm latency slightly increases. This is because DPA's memory subsystem and single-thread computing power are much wimpier than the host/Arm. These two experiments demonstrate the fragility of DPA's latency advantages, which can easily be surpassed in complex processing due to the host/Arm's stronger computing power and memory subsystem. 

\noindent{\bf Takeaways (Networking Latency).}
DPA is closer to the network compared with the host/Arm and thus has the lowest network latency. \sideword{R1}\zheng{However, this latency advantage is fragile and will be nullified when handling complex operations.}

\subsection{Network Throughput}
\label{section_net_bandwidth}
\begin{figure}[]
    \subfloat[Model=Send, payload=64B]{
        \label{fig_e_net_bandwidth_send_64B_threads}
        \includegraphics[keepaspectratio=true, width=0.48\linewidth]{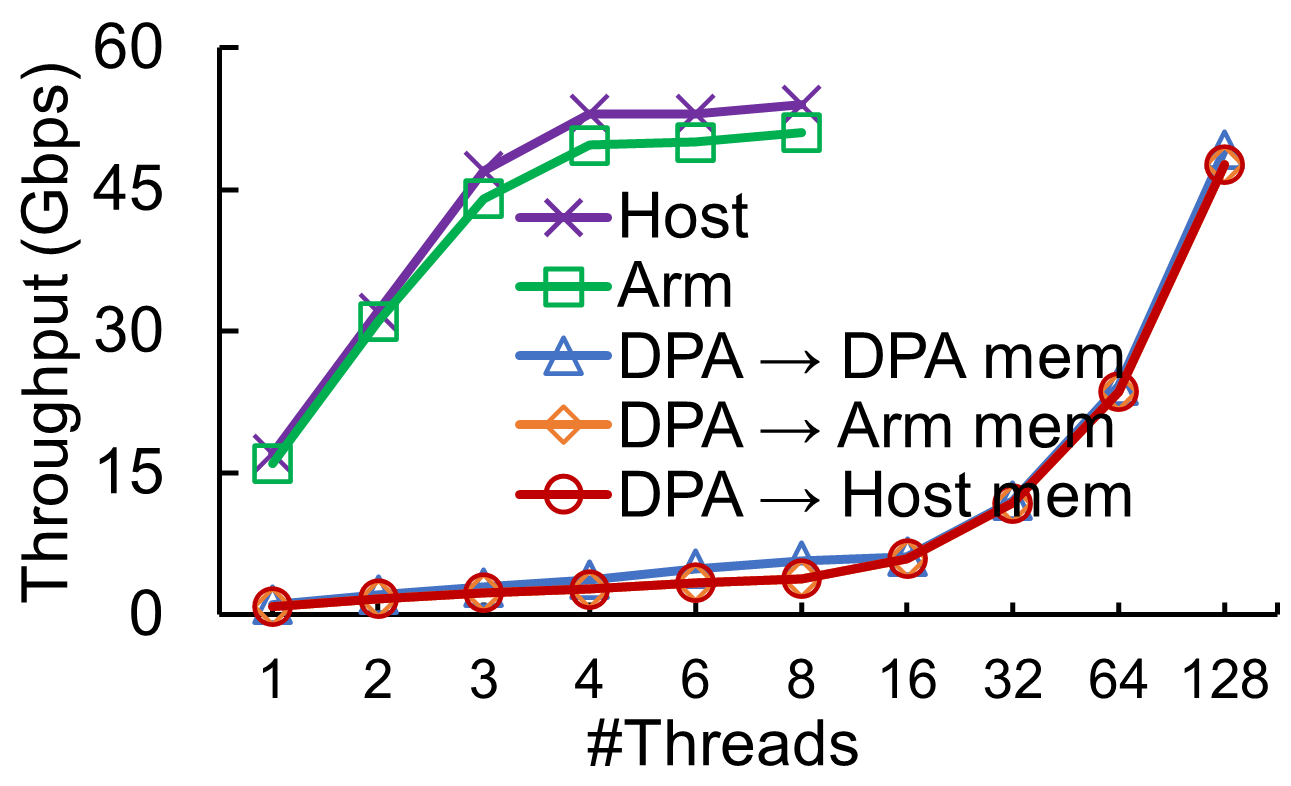}
    }
    \hspace{-2ex}
    \subfloat[Model=Receive, payload=64B]{
        \label{fig_e_net_bandwidth_receive_64B_threads}
        \includegraphics[keepaspectratio=true, width=0.48\linewidth]{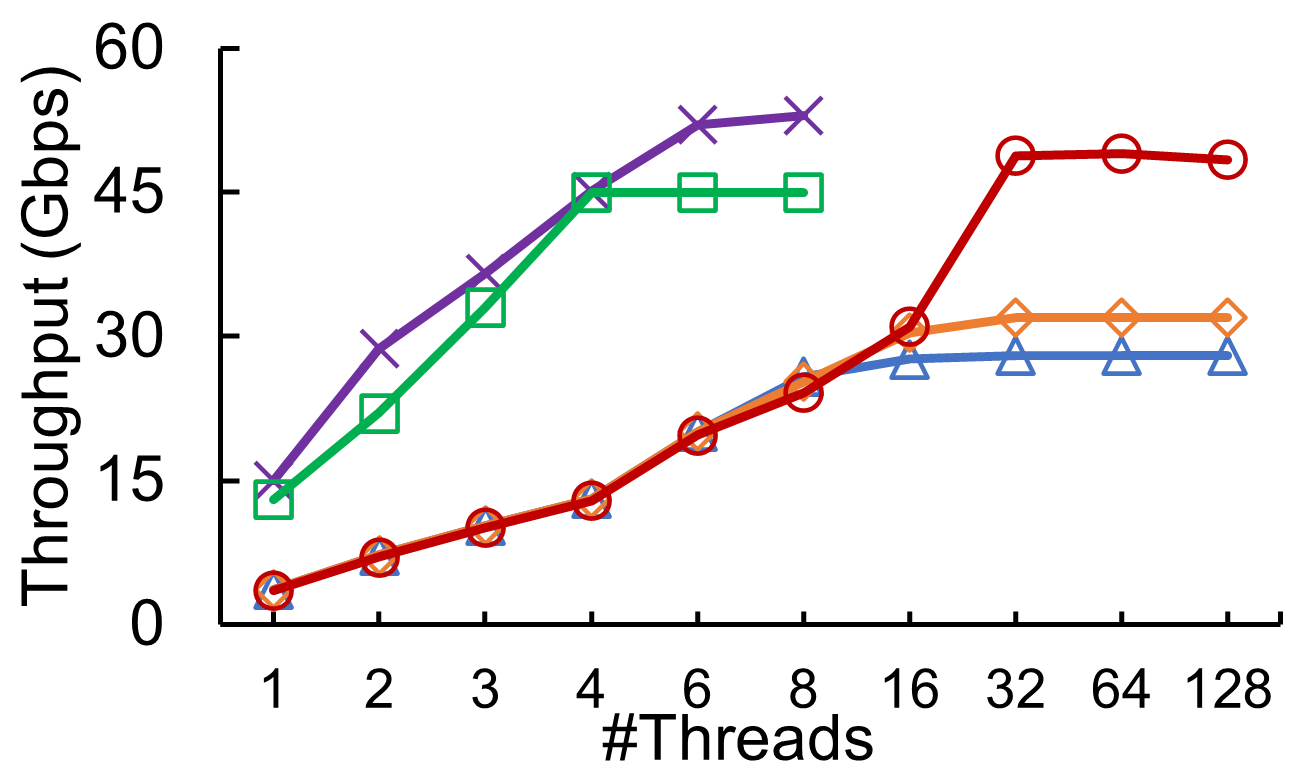}
    }
    \newline
    \subfloat[Model=Send, payload=1KB]{
        \label{fig_e_net_bandwidth_send_1k_threads}
        \includegraphics[keepaspectratio=true, width=0.48\linewidth]{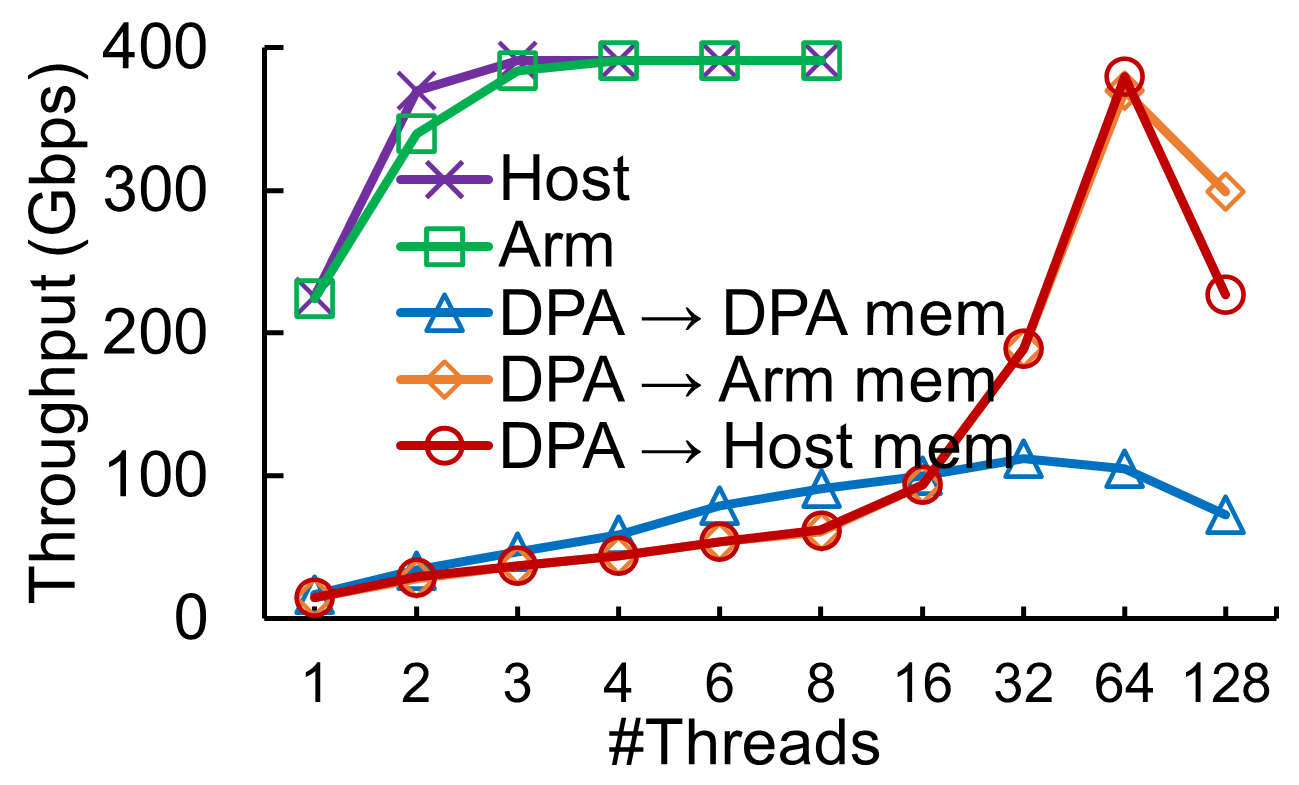}
    }
    \hspace{-2ex}
    \subfloat[Model=Receive, payload=1KB]{
        \label{fig_e_net_bandwidth_receive_1k_threads}
        \includegraphics[keepaspectratio=true, width=0.48\linewidth]{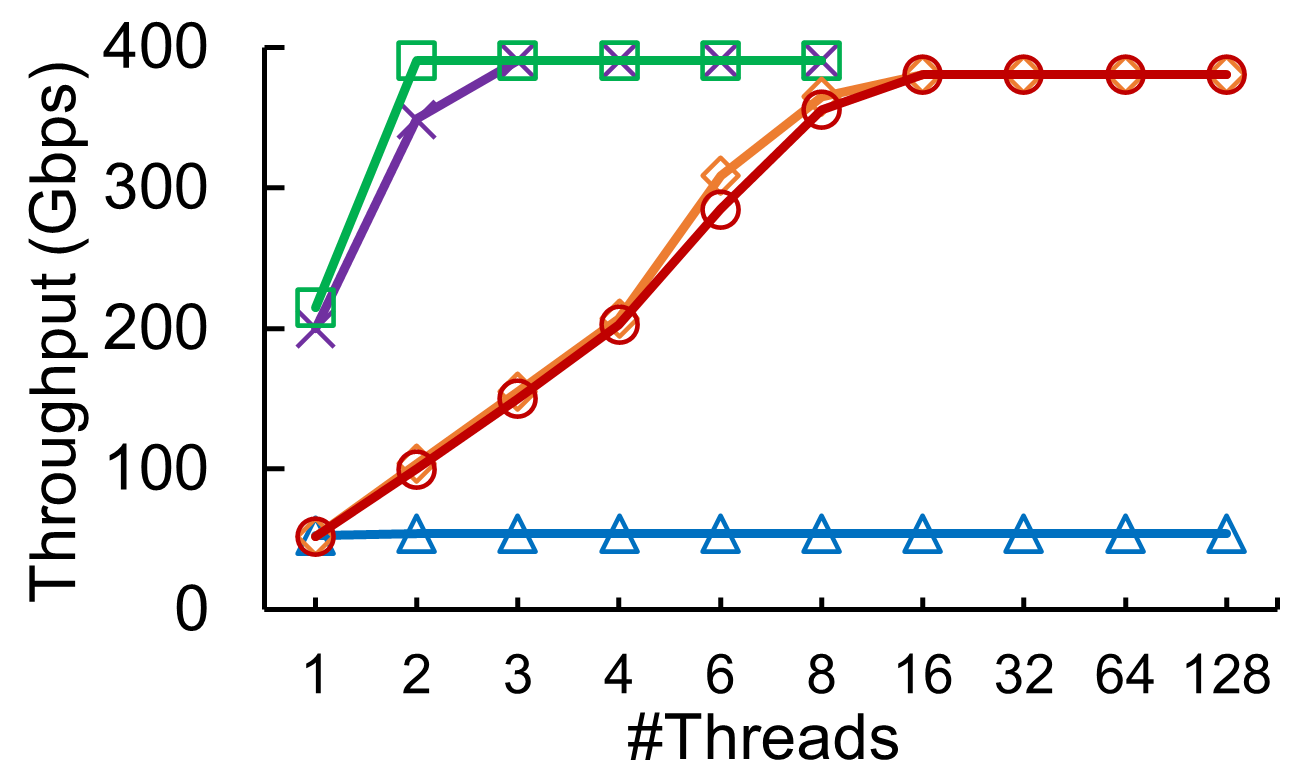}
    }
    \caption{Throughput comparison under different threads.}
    \label{fig_e_net_bandwidth_threads}
    \vspace{-3ex}
\end{figure}

In this subsection, we measure the achievable network throughput regarding different patterns (send/receive). For the DPA core, we use different memories as packet buffers. Figure~\ref{fig_e_net_bandwidth_threads} shows the achievable network throughput for different patterns under different packet sizes (64 Bytes and 1 KB). We have three observations. 

First, when the packet size is small (64 Bytes), none of the three processors can achieve the network line rate. DPA cores require more threads to achieve comparable throughput than the host CPU and Arm.

Second, when the packet size is large (1 KB), all three processors can achieve the network line rate. But DPA still requires more threads than the host CPU and Arm, indicating that the DPA thread is much wimpier than the host/Arm thread. Luckily, the number of DPA threads (256) is noticeably more than the Arm (16) and the host (32). So DPA cores can still achieve comparable network throughput using more threads. 

Third, when the packet size is large (1 KB), DPA must use Arm memory or host memory to achieve the network line rate. Using DPA memory only achieves around 100 Gbps for sending packets and around 50 Gbps for receiving packets. We speculate that this is due to the internal limitation of the DPA's L2/L3 cache since using host/Arm memory would bypass DPA's L2/L3 cache.

\noindent{\bf Takeaways (Networking Throughput).}
DPA thread is significantly wimpier than host/Arm threads regarding sending/receiving network packets, but DPA has more threads and can achieve comparable network throughput as host/Arm. 

\noindent{\bf DPA’s unique networking characteristic:}
DPA has the advantage of being closer to the network and thus enjoys lower network latency. Programmers can consider offloading latency-sensitive network applications to the DPA. However, DPA's single-thread performance is too wimpy, so the offloaded workloads can not involve heavy computation or many memory operations. Otherwise, the latency advantages would quickly be overshadowed. We further use a case study to measure the impact of DPA’s unique network characteristic on the end-to-end applications in~\S~\ref{section_case_study_ntp}.

\vspace{-1ex}
\section{Case Studies}
\label{sec_case_study}
In this section, guided by the above benchmarking results, we further use three case studies to explore the three unique characteristics of the DPA cores. \sideword{R3}\zheng{In each case study, we have five kinds of implementations unless stated otherwise.

\squishlist
\item ``Host": all functions are deployed in the host CPU.
\item ``Arm": all functions are deployed in Arm.
\item ``DPA$\rightarrow$Y mem": all functions are deployed in DPA. ``Y mem" indicates the network buffer memory type (host/Arm/DPA memory) used.
\squishend
}

\subsection{Clock Synchronization Service}
\label{section_case_study_ntp}
Compared with the host/Arm, DPA is the closest to the network. Latency-sensitive network workloads can benefit from DPA's characteristic that is closer to the network. 

To explore the potential impact, we conduct a case study on latency-sensitive clock synchronization service. Clock synchronization is critical for many datacenter applications~\cite{sundial_osdi20} such as distributed transactional databases~\cite{spanner_tocs13,k_database}, consistent snapshots~\cite{k_snapshot_0,k_snapshot_1}, and network telemetry~\cite{k_telemetry_0, k_telemetry_1}. The key metric for clock synchronization is the $time~uncertainty~bound$ for each node~\cite{spanner_tocs13, sundial_osdi20, k_time_uncertainty_bound}, donated as $\epsilon$. Lower time uncertainty bound ($\epsilon$) can improve the performance of distributed applications and enable more accurate one-way delay measurement. 

\noindent{\bf Implementations.}
We use two servers mentioned in~\S~\ref{sec_server_setup}, one as a synchronization client and the other as a synchronization master node. The synchronization duration is set to be 0.1 seconds and the clock drift is set to be 10 microseconds per second~\cite{k_clock_drift}. We have deployed five kinds of clock synchronization services as mentioned above.

\begin{figure}[]
    \subfloat[Without network pressure]{
        \label{fig_e_ntp_avg_tub}
        \includegraphics[keepaspectratio=true, width=0.50\linewidth]{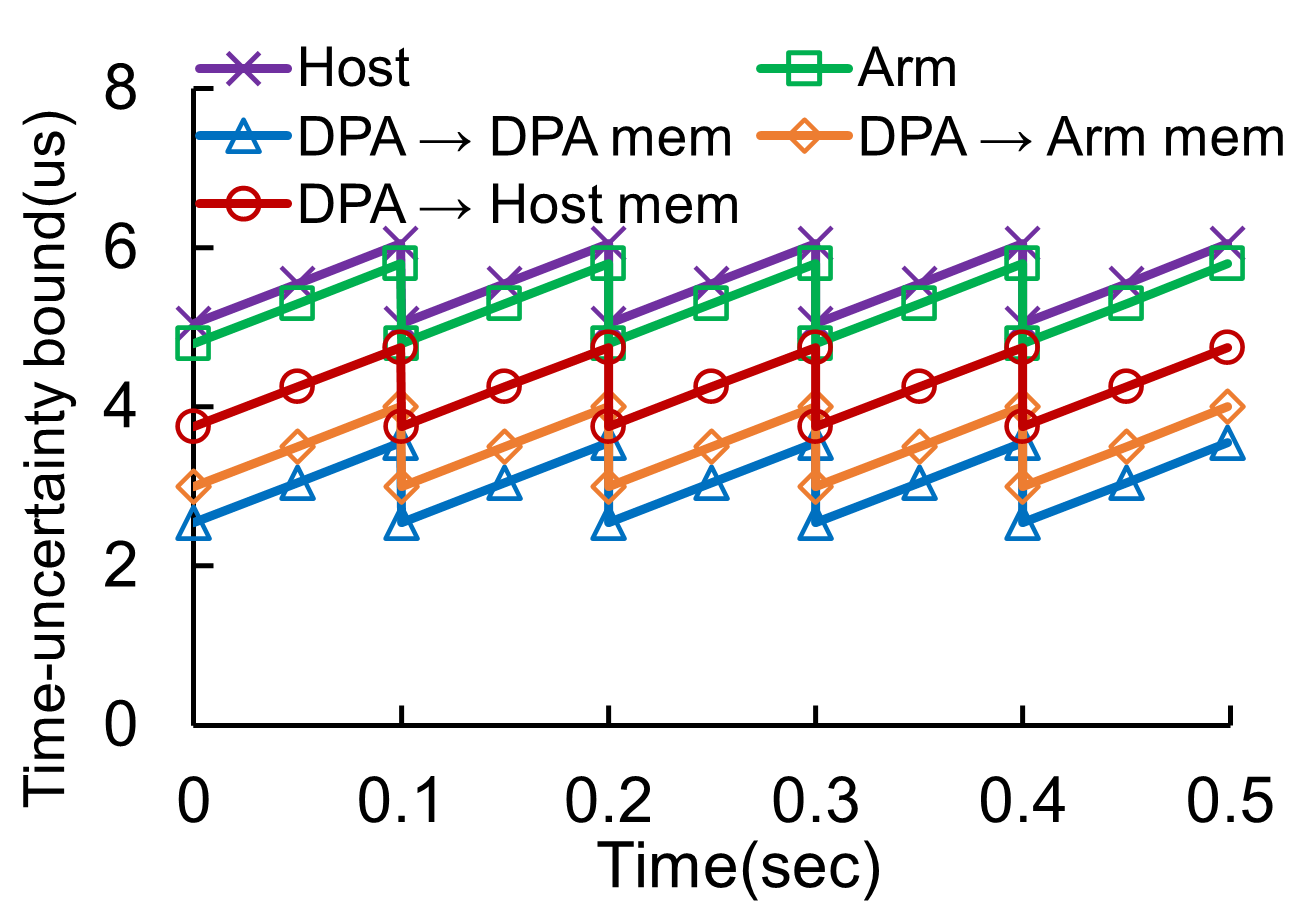}
    }
    \subfloat[P999 under 400Gbps network]{
        \label{fig_e_ntp_p999_tub}
        \includegraphics[keepaspectratio=true, width=0.50\linewidth]{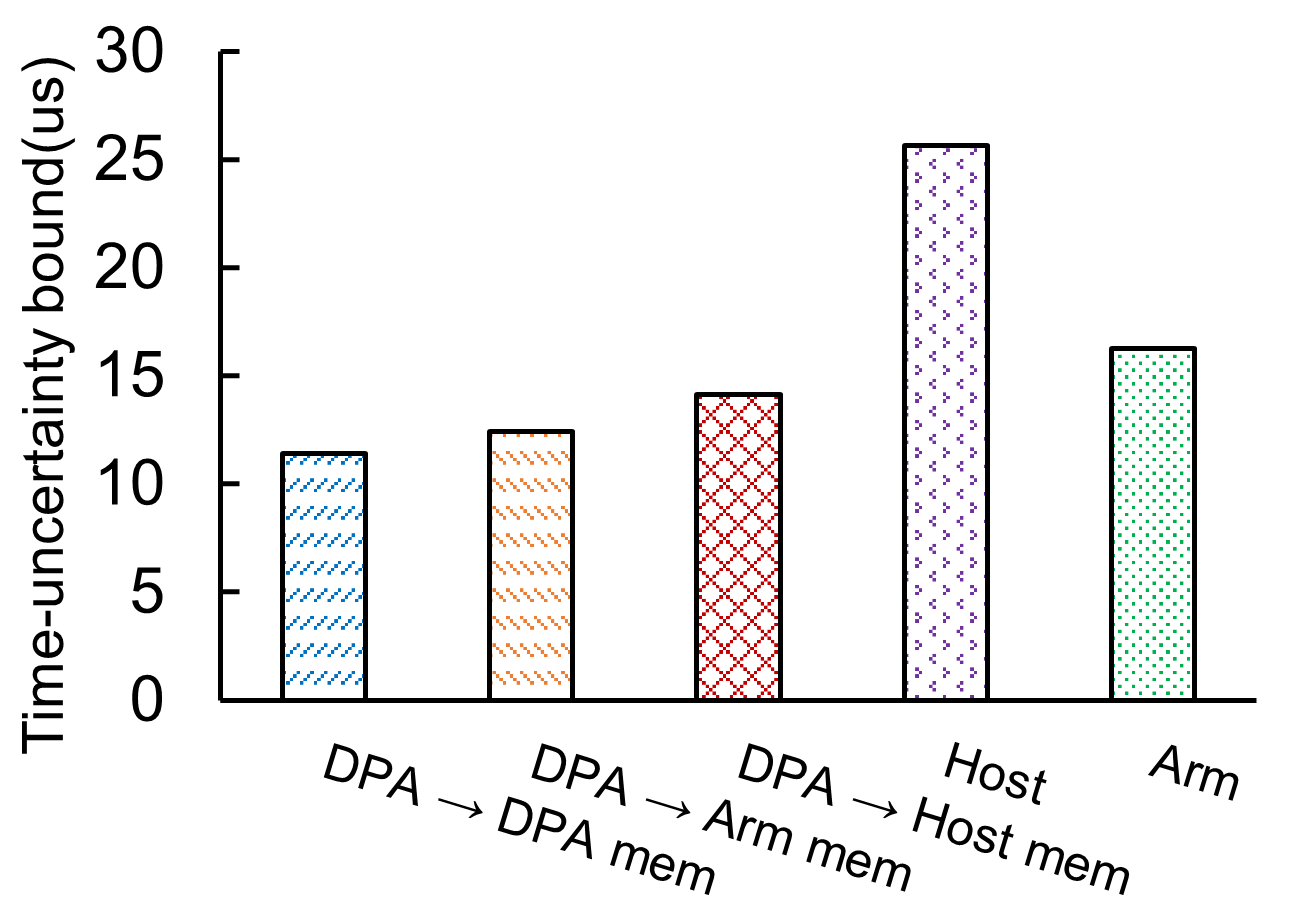}
    }
    \caption{Time-uncertainty bound under different patterns.} 
    \label{fig_e_ntp}
    \vspace{-2ex}
\end{figure}

Figure~\ref{fig_e_ntp_avg_tub} shows the time uncertainty bound ($\epsilon$) for different implementations when the network is under-loaded, i.e., there is only a clock synchronization service in each server. Figure~\ref{fig_e_ntp_p999_tub} shows the 999th percentile time uncertainty bound when the network is heavily used. To make the network heavily used, we let an extra network-intensive L2-reflector application run in the host cores at the same time and it consumes up to 400 Gbps bi-direction network throughput. We have two observations. 

First, all three DPA implementations offer much lower time uncertainty bound than the host/Arm, up to 2.0$\times$ in average latency and 2.3$\times$ in 999th percentile latency. This indicates that DPA's characteristics can greatly improve the performance of latency-sensitive network applications due to its location closer to the network. Second, among three DPA implementations, ``DPA$\rightarrow$DPA mem'' offers the lowest time uncertainty bound. This is because the network packets can directly be put into or be fetched from DPA's L2/L3 cache when DPA cores are using DPA memory. This indicates that latency-sensitive network applications running in the DPA cores should use DPA memory as the network packet buffer. 

\noindent{\bf Guideline 1: Offloading latency-sensitive logic to DPA to improve performance.} DPA is closer to the network compared with the host/Arm. An application that is sensitive to network latency and fits within DPA cache working set size can leverage this characteristic to improve end-to-end performance. Also, the programmer should choose DPA memory as the network buffer to promote incoming network packets to DPA caches and thus further decrease the processing latency. 

\subsection{Network Function Virtualization}
\label{section_case_study_nf}
Compared with the host/Arm, DPA has a thread number that is an order of magnitude higher. To fully enjoy the vast number of threads, the offloaded workloads should be very easy to parallelize, otherwise, the application performance can significantly downgrade due to DPA's poor single-thread computing ability.

In this case study, we leverage stateless network function virtualization (NFV)~\cite{k_nfv, microboxes, hypernf} to explore DPA's many-core parallelism. Stateless network functions are usually easy to parallelize and can scale well with the number of threads.

\noindent{\bf Implementations.} We focus on the throughput metric instead of the latency metric. We choose two stateless network functions: L2-reflector and CheckIPHeader. These two network functions are not memory-intensive such that we can minimize the impact of DPA's memory subsystem characteristics. We have five kinds of implementations as mentioned above.

\begin{figure}[]
    \subfloat[NF=L2 reflector, payload=64B]{
        \label{fig_e_nf_l2_reflector_64B}
        \includegraphics[keepaspectratio=true, width=0.50\linewidth]{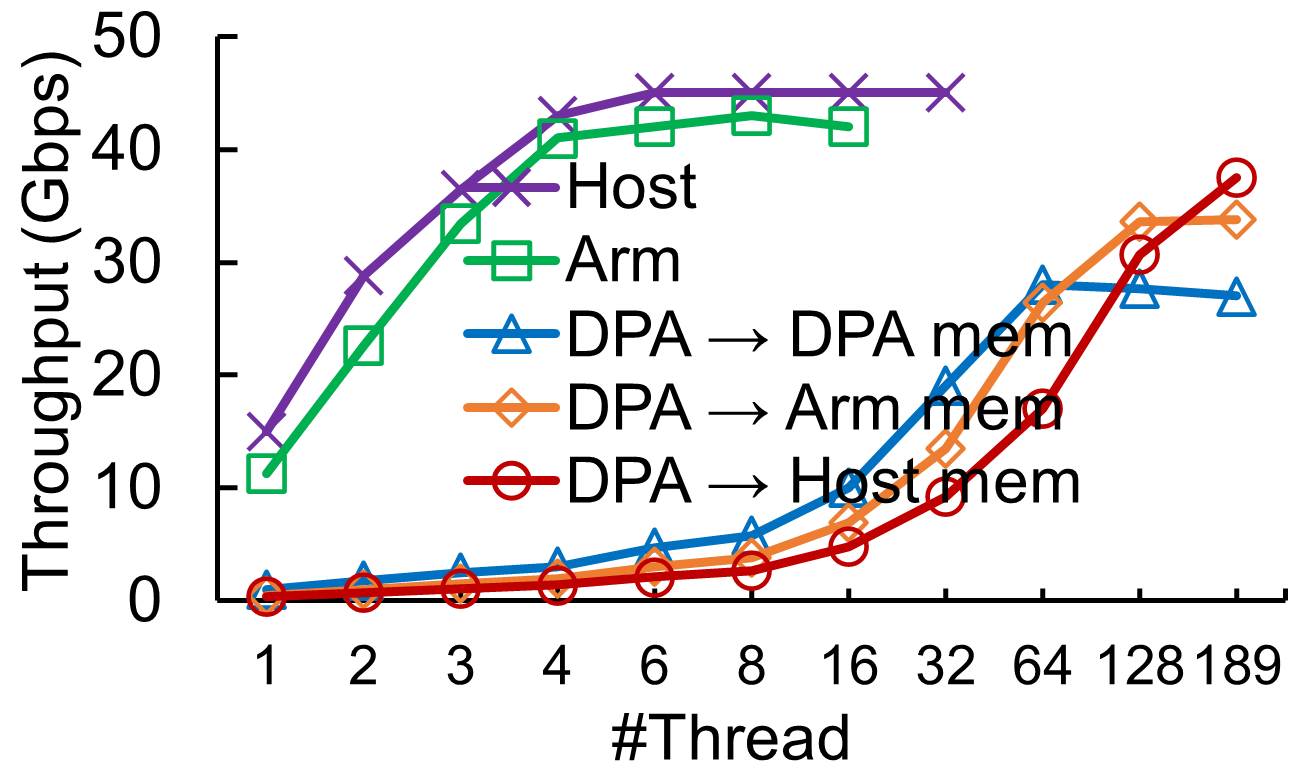}
    }
    \subfloat[NF=IP check header, payload=64B]{
        \label{fig_e_nf_udp_check_header_64B}
        \includegraphics[keepaspectratio=true, width=0.50\linewidth]{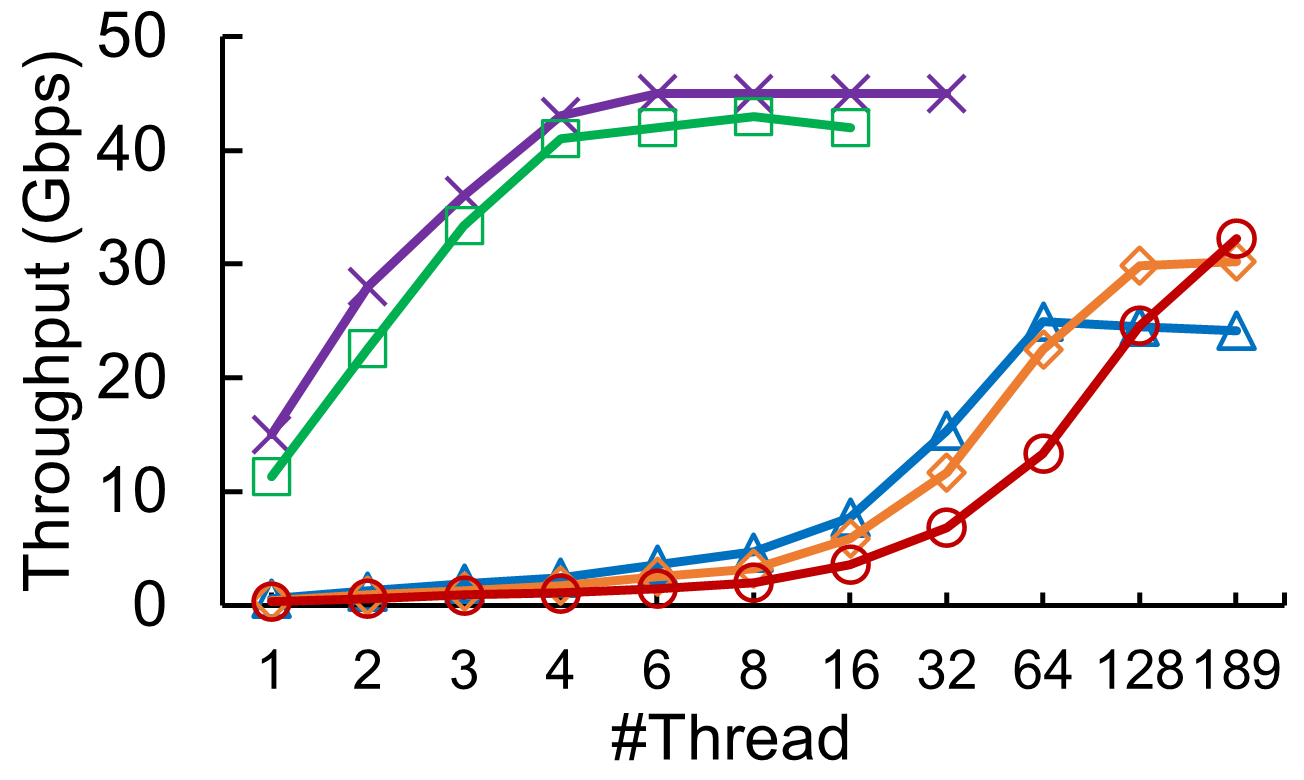}
    }
    \vspace{-1ex}
    \newline
    \subfloat[NF=L2 reflector, payload=1024B]{
        \label{fig_e_nf_l2_reflector_1k}
        \includegraphics[keepaspectratio=true, width=0.50\linewidth]{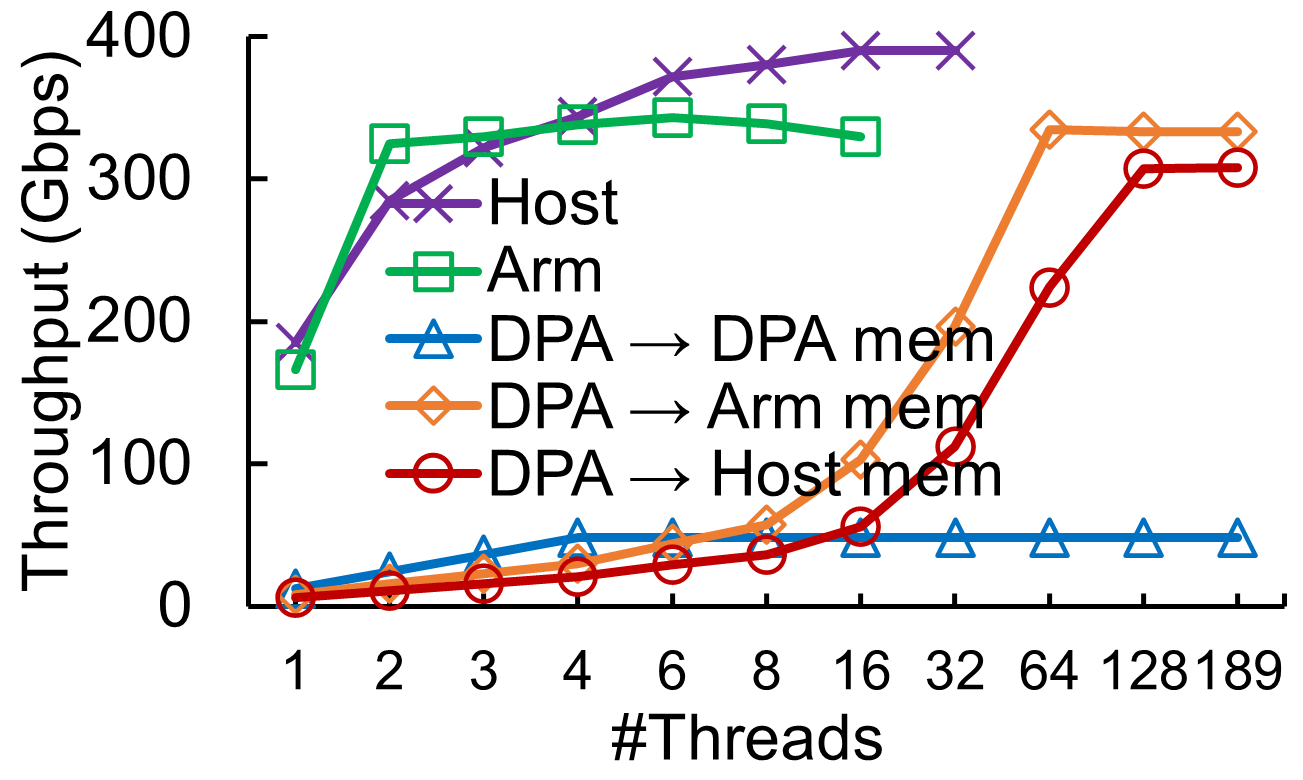}
    }
    \subfloat[NF=IP check header, payload=1024B]{
        \label{fig_e_nf_udp_check_header_1k}
        \includegraphics[keepaspectratio=true, width=0.50\linewidth]{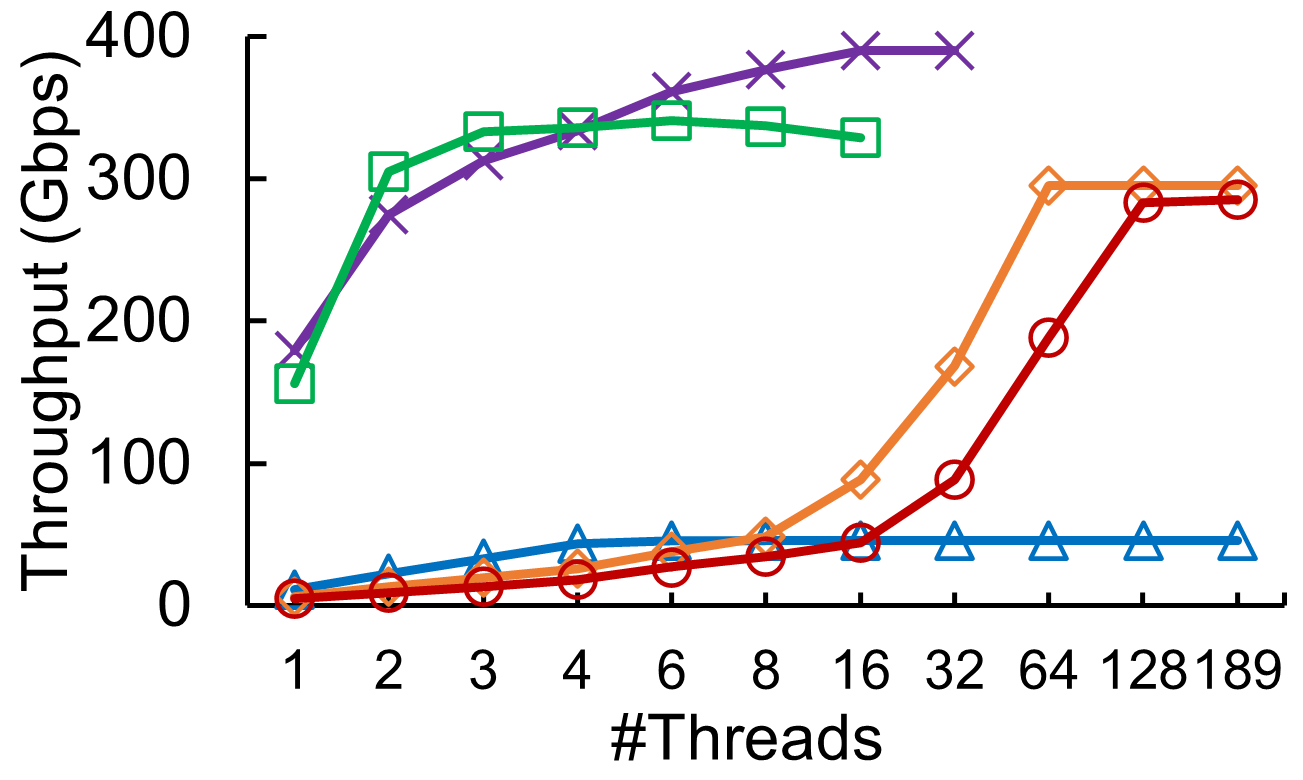}
    }    
    \caption{L2 reflector and IP check header network functions under different network patterns.}
    \label{fig_e_network_function}
    \vspace{-3ex}
\end{figure}

Figure~\ref{fig_e_network_function} shows the result when packet size is 64B and 1KB. We have two observations. First, DPA's single thread throughput is substantially lower than that of the host/Arm, which is within our expectations. However, with its high thread count, DPA can still achieve comparable throughput as the host/Arm. This indicates that programmers should consider offloading easy-to-parallelize workloads to DPA cores, otherwise, the system throughput may drop significantly (up to several magnitudes) due to DPA's wimpy single-thread performance. Second, when the packet size is large (1KB), the throughput of ``DPA$\rightarrow$DPA mem'' can not scale along the number of threads. This is consistent with our results from previous network throughput measurements~\S~\ref{section_net_bandwidth}, i.e., the maximum send and receive throughput of ``DPA$\rightarrow$DPA mem'' is around 100 Gbps and 50 Gbps, respectively. This indicates that throughput-intensive applications running in the DPA should avoid using DPA memory as the network buffer.

\noindent{\bf Guideline 2: Offload easy-to-parallelize applications to DPA.} \sideword{R1}\zheng{Although DPA's single thread is very wimpy, in terms of computing power and memory bandwidth, its thread count (256) is far more than the host/Arm, and can provide line-rate processing ability while processing stateless network function workloads. } We suggest programmers can consider using easy-to-parallelize workloads to leverage DPA's many-core characteristics. \zheng{In addition, using DPA memory for memory-bandwidth-intensive applications is not the best solution due to internal DPA cache limitation.}


\subsection{Key-value Aggregation}
\label{section_case_study_kv_aggr}

To explore the potential impact of this characteristic, we conduct a case study on memory-intensive key-value stream aggregation~\cite{ask_asplos23}. Key-value stream aggregation is a memory-intensive operation widely existing in various distributed systems, e.g., $reduce()$ in big data processing~\cite{mapreduce,k_big_data}, $AllReduce()$ in distributed training~\cite{k_dist_ml_0,k_dist_ml_1,k_dist_ml_2, switchml}, $MPI\_Reduce()$ in high-performance computing~\cite{k_hpc}, etc. Key-value aggregation is considered to be memory-intensive because the application needs a large memory to hold the intermediate aggregation results and needs to frequently update the intermediate results according to the received network packets. We explore DPA's memory characterization by carefully optimizing the DPA implementation.

\begin{figure}[]
    \subfloat[$2^{16}$ distinct keys]{
        \label{fig_e_kv_aggregation_uniform_tuple}
        \includegraphics[keepaspectratio=true, width=0.5\linewidth]{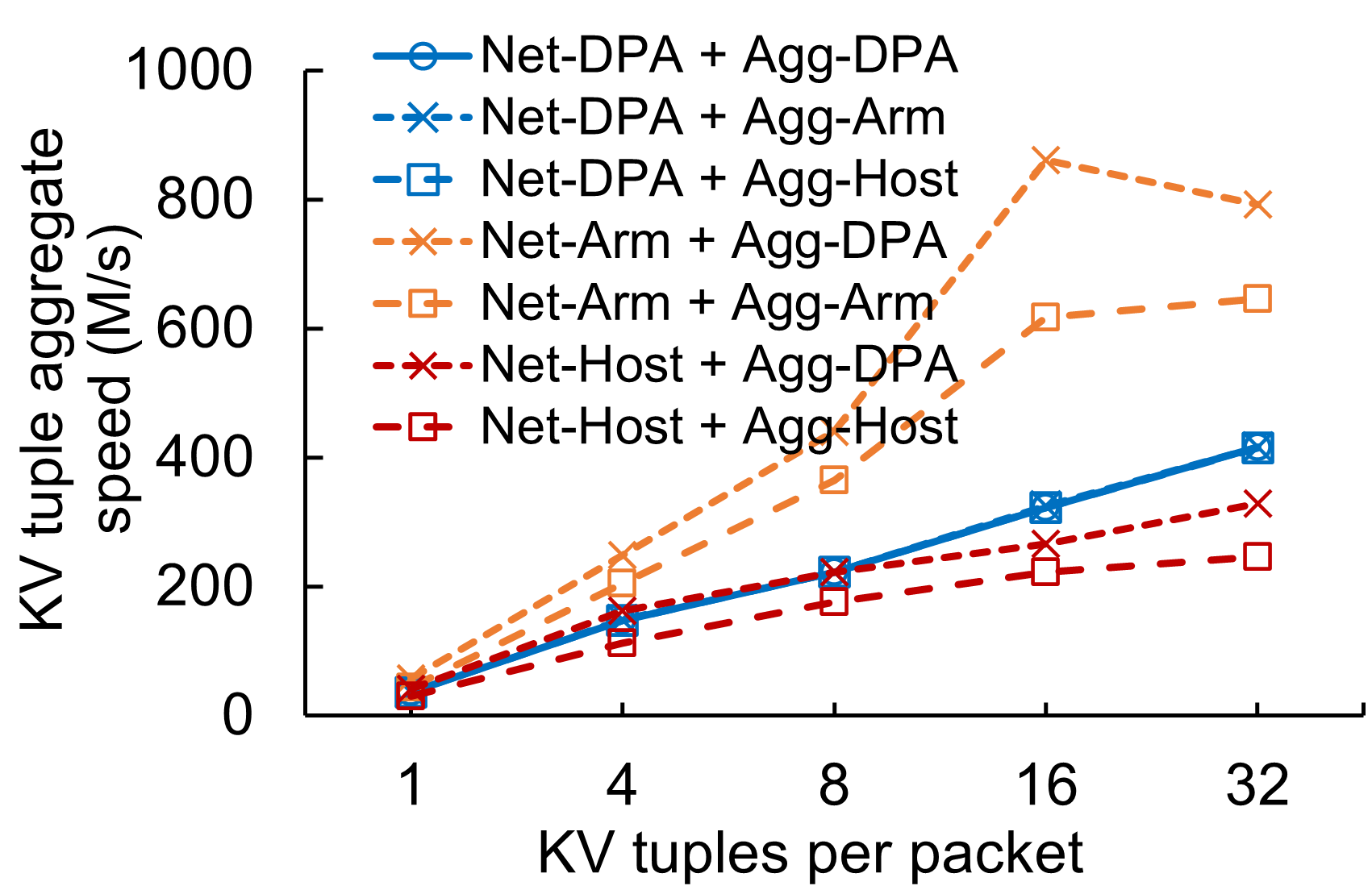}
    }
    \subfloat[32 KV tuples per packet]{
        \label{fig_e_kv_aggregation_uniform_keys}
        \includegraphics[keepaspectratio=true, width=0.5\linewidth]{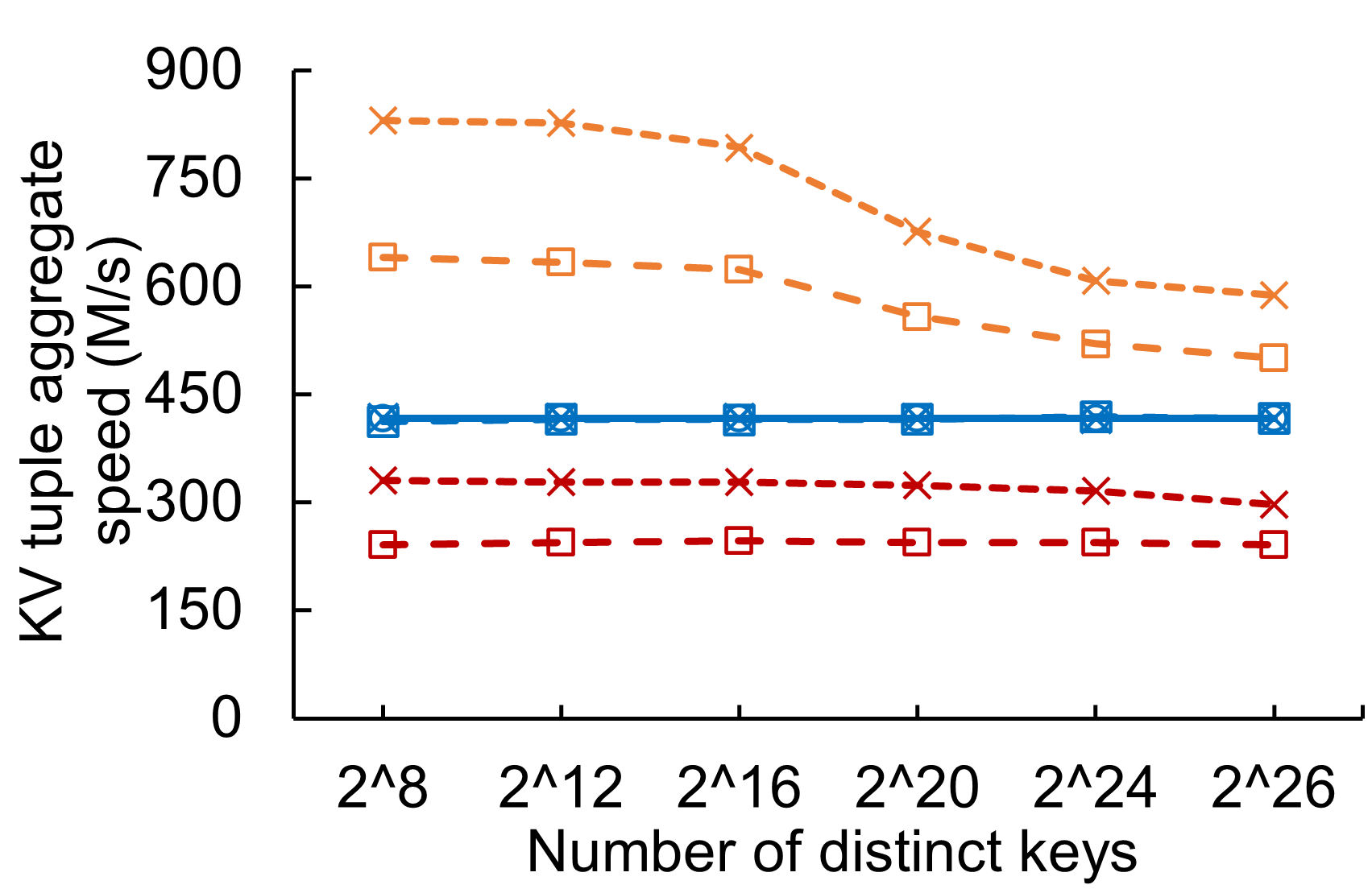}
    }
    \caption{Effect of memory types on DPA's throughput.} 
    \label{fig_e_kv_aggregation_custom_uniform_dataset}
    \vspace{-3ex}
\end{figure}

\begin{figure}[]
    \vspace{-1ex}
    \subfloat[1 KV tuple per packet]{
        \label{fig_e_kv_aggregation_yelp_1tuple}
        \includegraphics[keepaspectratio=true, width=0.5\linewidth]{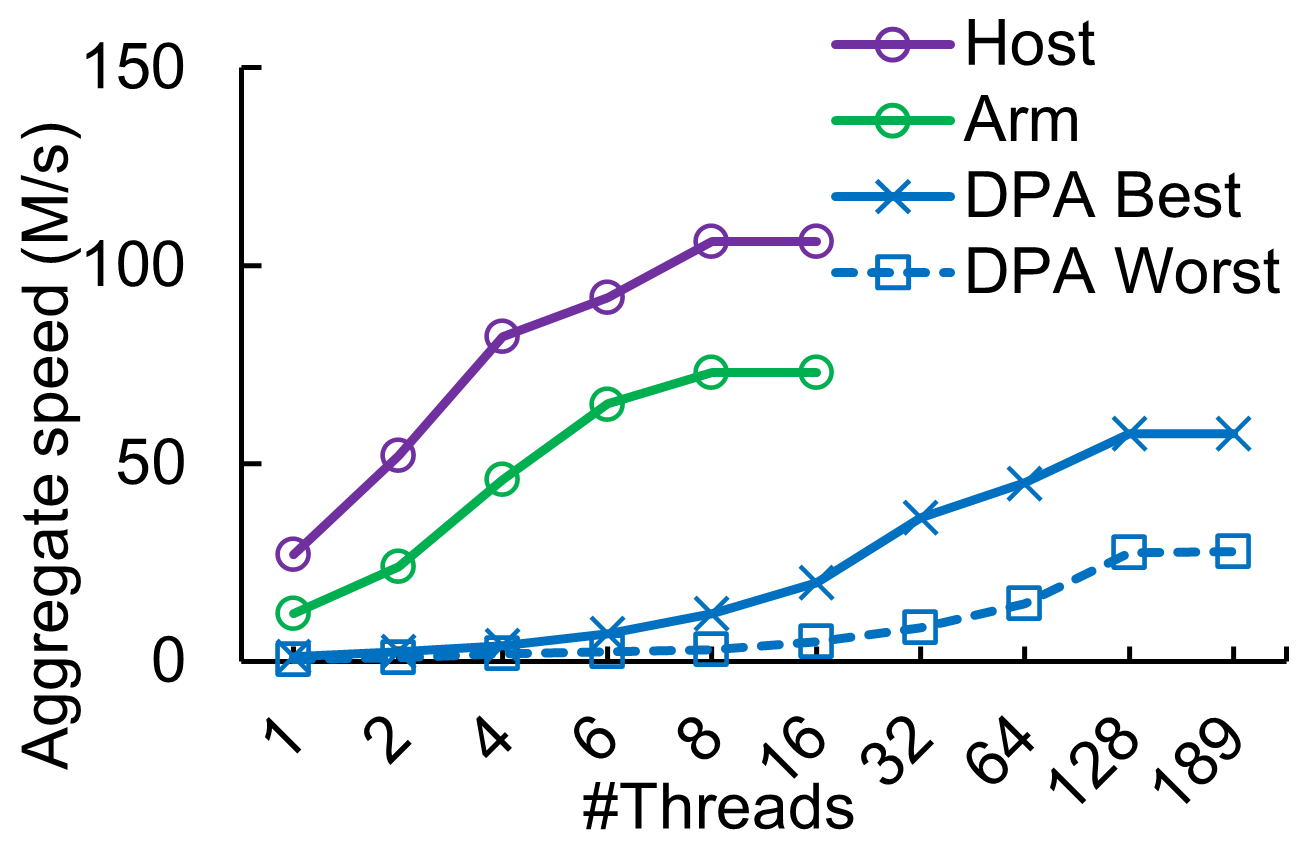}
    }
    \subfloat[8 KV tuples per packet]{
        \label{fig_e_kv_aggregation_yelp_8tuple}
        \includegraphics[keepaspectratio=true, width=0.5\linewidth]{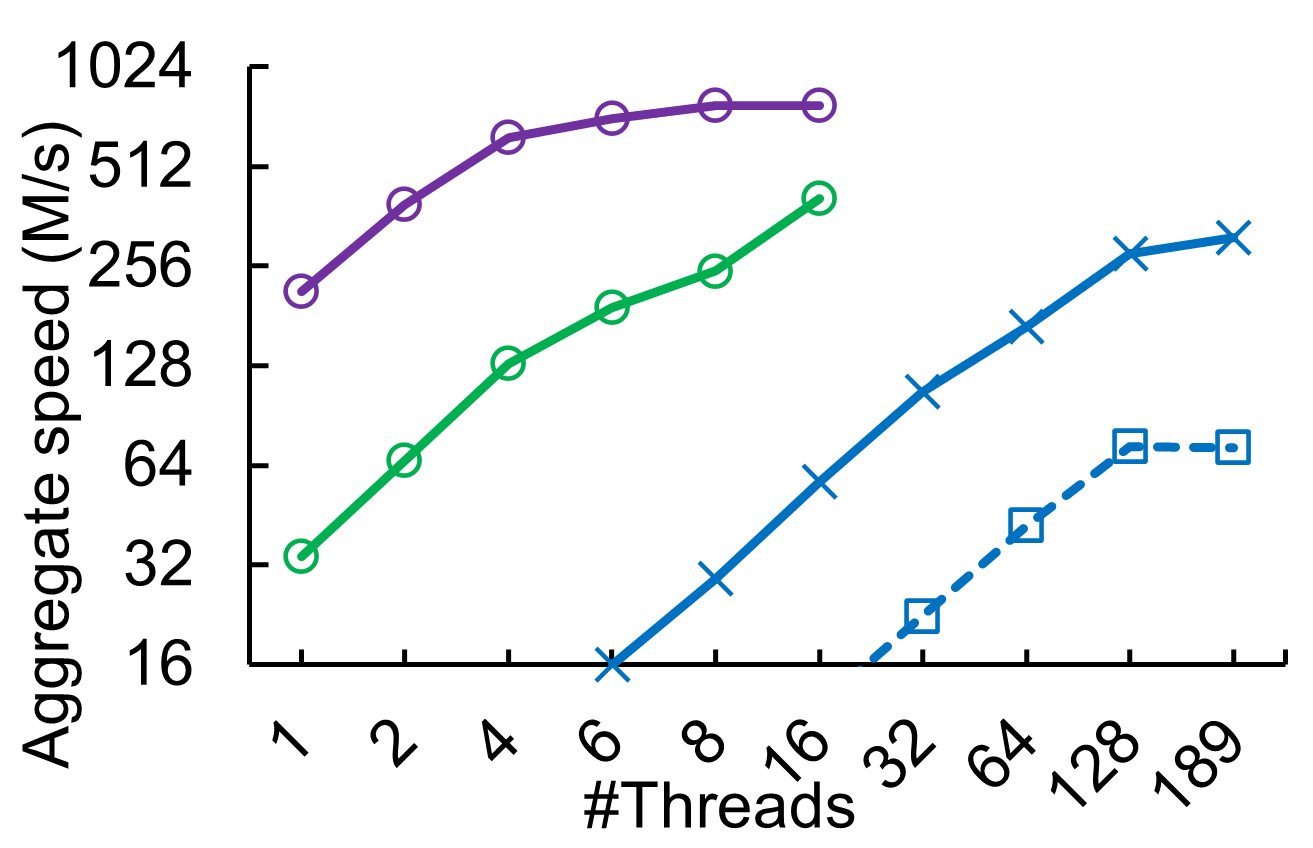}
    }
    \caption{Throughput comparison between three processors.} 
    \label{fig_e_kv_aggregation_yelp_dataset}
    \vspace{-3ex}
\end{figure}

\noindent{\bf Implementations.} When running key-value stream aggregation, we mainly have two memory regions: network buffer (NetBuf) and aggregation buffer (AggBuf). The network buffer is used to send/receive network packets (also known as network data ring buffer) while the aggregation buffer is used to hold the intermediate aggregation results for each key. \sideword{R4}\zheng{We have seven implementations with supplementary explanations.}

\squishlist
\item ``Net-X+Agg-Y'': the aggregation service is deployed in the DPA cores. ``X" indicates the used memory type for NetBuf while ``Y" indicates the used memory type for AggBuf.
\squishend

\noindent{\bf Effect of Memory Type.} We first evaluate the effect of memory types on the DPA's end-to-end key-value aggregation throughput. We generate an artificial uniform distribution trace to understand the effectiveness of hot-key. The key-value tuple size is 16 bytes and each packet has a 64 bytes header. 

Figure~\ref{fig_e_kv_aggregation_uniform_tuple} shows the achieved aggregation throughput under different key-value tuples per packet. We observe that ``Net-Arm+Agg-DPA'' always achieves the highest throughput. Regarding NetBuf, we should use the host memory or Arm memory, as using DPA memory to handle network packets has up to 7.5$\times$ lower throughput than the host/Arm (\S~\ref{section_net_bandwidth}). Regarding AggBuf, we should consider using the DPA memory or Arm memory for their higher memory bandwidth than the host (\S~\ref{section_memory_bandwidth}). Among these possible combinations, ``Net-Arm+Agg-DPA'' provides the highest throughput. 

Figure~\ref{fig_e_kv_aggregation_uniform_keys} shows the aggregation throughput under different numbers of distinct keys. The per-packet KV tuples are fixed to 32. We have three observations. 
First, when the number of distinct keys increases, the throughput for ``Net-Arm+Agg-DPA'' and ``Net-Arm+Agg-Arm'' noticeably decreases because the number of distinct keys exceeds the DPA's L2/L3 cache size and Arm's L3 cache size. 

Second, all implementations that use DPA memory as NetBuf (all blue lines) suffer from low throughput since they are bounded by DPA memory's limited ability to receive packets. 

Third, two implementations that use the host memory as NetBuf (two red lines) show the lowest throughput because the Key-value aggregation service requires reading the received packets after receiving the packets into the memory, while the throughput of the DPA core accessing the host memory is very low (7.2 GB/s read bandwidth and 14 GB/s write bandwidth).

\noindent{\bf Comparisons with the host/Arm.}
We then compare the aggregation throughput of the host/Arm/DPA. We use ``yelp'' trace from production~\cite{yelp_dataset}. We select the best(``Net-Arm+Agg-DPA'') and worst(``Net-Host+Agg-Host'') results from all combinations for DPA (named ``DPA-Best'' and ``DPA-Worst'').
Figure~\ref{fig_e_kv_aggregation_yelp_dataset} shows the achieved aggregation throughput under different thread numbers. We have two observations. 

First, DPA's best-case aggregation throughput is lower but comparable to the host (2.5$\times$)/Arm (1.3$\times$). Although lower than the host/Arm, DPA's end-to-end aggregation throughput still surprises us, considering that DPA's memory latency is an order of magnitude higher than the host/Arm and the memory throughput is significantly lower than the host/Arm.

Second, DPA's best-case aggregation throughput is markedly up to 4.3$\times$ higher than that of the worst-case. This indicates the importance of selecting appropriate memories for different usages when running applications in DPA cores.

\noindent{\bf Guideline 3: DPA programmers should carefully choose memories from the host/DPA/Arm memory.} Using different memories for DPA can result in significantly different performance. \sideword{R1}\zheng{The performance gap between the best (Net-Arm + Agg-DPA) and worst (Net-Host + Agg-Host) combinations can be as high as several times.} Programmers must choose different memories according to the specific usage, to guarantee high performance when implementing on DPA.

\vspace{-1ex}
\section{Discussion}
\noindent{\bf Conclusion of guidelines.} In this section, we conclude three important DPA-related guidelines for future datapath accelerator-enhanced SmartNIC programmers, followed by our suggestions for DPA-enhanced SmartNIC vendors. 

\squishlist
\item \textbf{1. Offloading latency-sensitive and simple workloads to DPA cores.}
DPA is much closer to the network and enjoys lower network latency. Latency-sensitive applications can exploit this characteristic to improve end-to-end performance. Meanwhile, DPA's single-thread performance is minimal and the offloaded applications cannot involve too heavy computing, otherwise, the low-latency benefit can be quickly overshadowed.

\item \textbf{2. Offloading easy-to-parallelize workloads with small working set sizes.}
Our benchmarking results reveal that DPA has more cores than the host/Arm, while each thread is much wimpier. Programmers can offload easy-to-parallelize workloads to exploit DPA's many-core parallelism. However, when the application's working set size exceeds DPA's cache size, DPA has to frequently access memory and thus suffers from severe performance downgradation. Therefore, the memory working set size should better fit in DPA's cache size to avoid significant performance downgradation. 

\begin{figure}
	\centering
	{\includegraphics[keepaspectratio=true, width=0.9\linewidth]{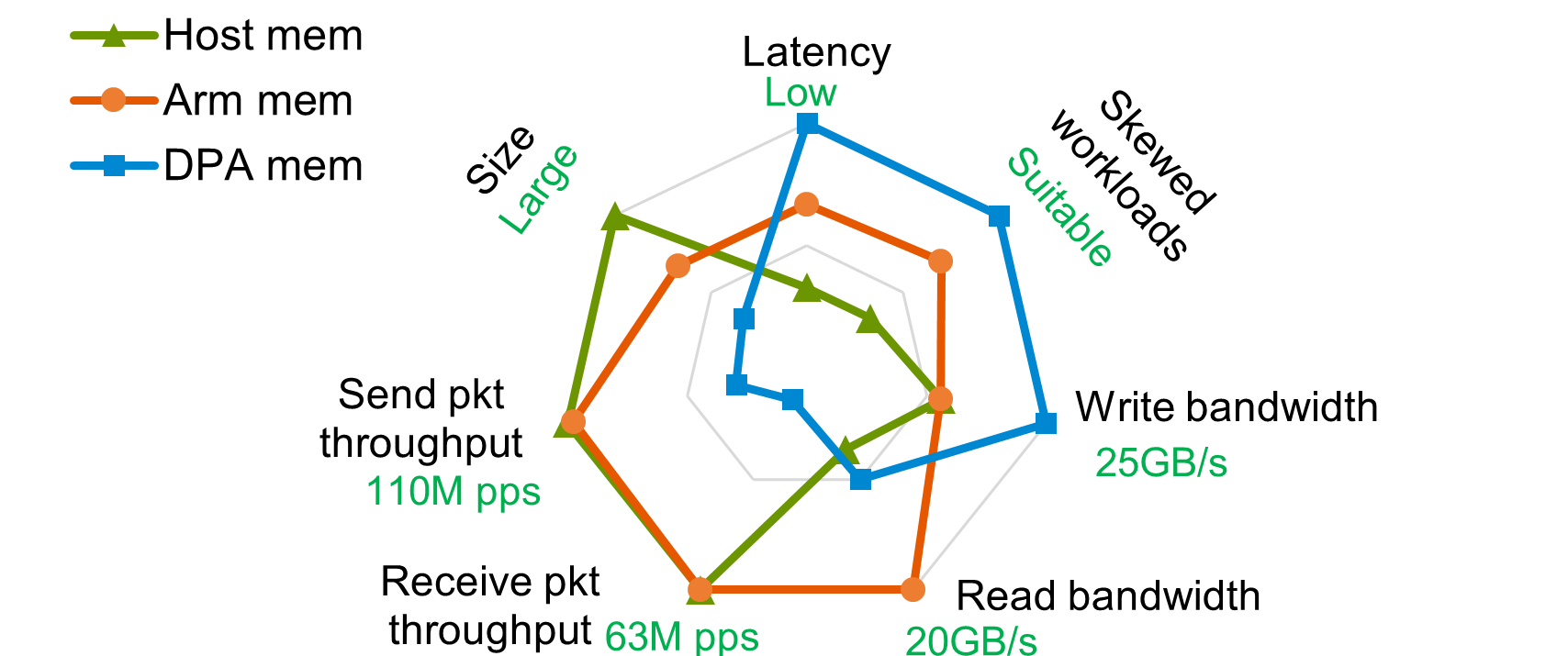}
	\caption{Radar chart for DPA under three memory types. The more prominent a line appears on a radar chart, the more indicative it is of a better performance.}
	\label{fig_e_kv_aggregation_radar}}
    \vspace{-4ex}
\end{figure}

\item \textbf{3. Carefully select memory buffers for DPA applications.}
DPA can not only access its own memory/cache but also access the off-path Arm's memory and the host CPU's memory. Placing application buffers in any of three kinds of memory leads to significantly different performance. Therefore, we list seven main indicators that programmers may be concerned about when choosing memories, as shown in Figure~\ref{fig_e_kv_aggregation_radar}. where a larger value in each axis is better. We highlight three noticeable hints. 
First, use the host memory or the Arm memory as the network buffer for bandwidth-intensive network applications, since DPA memory performs poorly regarding the throughput of sending/receiving packets. 
Second, using host memory is more suitable for DPA applications with large memory footprints because its available size is substantially larger than DPA/Arm memory, given that the host X86 has up to eight per socket memory channels while BF3 only features two DDR5 channels. 
Third, DPA memory has the potential to perform better in skewed workloads since it features an additional DPA L2/L3 cache that is closer to the DPA core.
\squishend

\noindent{\bf Suggestions for SmartNIC vendors.} Our characterization has uncovered several important factors that bound the performance of the datapath accelerator. And we have some suggestions for SmartNIC vendors.

\squishlist
\item \textbf{1. Directly attach a memory to the DPA.} Currently, DPA has no directly attached memory, all memory accesses have to go through a high-latency NIC switch. This incurs at least 5$\times$ higher memory access latency than that of Arm. This can be harmful for applications whose working sets can not be fully cached by DPA's cache. As such, we suggest vendors consider directly attaching a memory to the DPA.

\item \textbf{2. Equip DPA with a more powerful cache.} Current DPA cache suffers from very high latency (up to 10.5$\times$ higher L1 latency than the host CPU cache). Besides, the per-thread L1 cache bandwidth is only 0.53 GB/s (up to 92$\times$ lower than the host L1 cache). Although the frequency of DPA (1.8 GHz) is not significantly lower than that of the host (2.5 GHz), DPA's uncore frequency is 4.4$\times$ lower than the host, which may explain why the DPA cache performance is so poor. As such, we suggest vendors consider equipping DPA with a more powerful cache.

\squishend

\vspace{-1ex}
\section{Related Works}
To our knowledge, this paper is the first to benchmark DPA-enhanced SmartNIC.

\noindent{\bf Characterizing SmartNICs.}
Prior works~\cite{ipipe,ipads_bf2,fudan_bf2,bf2_performance,battle_bf3_hoti,dpubench} characterize SmartNICs from different perspectives. iPipe~\cite{ipipe} characterizes SoC SmartNICs regarding their computing capacity, memory, traffic control, and host communication. In contrast, we characterize the different processors in a BF3-attached server from architectural perspectives and explore the possible influence of the unique architectural characteristics of the DPA. Wei et al.~\cite{ipads_bf2} characterize the off-path BlueField-2 SmartNIC from a communication-path perspective and concurrently use multiple communication paths to improve throughputs. In contrast, we focus on the architectural differences of processors within a BF3-attached server and optimize offloading performance according to the architectural characteristics. Sun et al.~\cite{fudan_bf2} evaluate off-path BlueField-2 using high-level applications and suggest using the SmartNIC as a new endpoint to provide horizontal scaling. Liu et al.~\cite{bf2_performance} characterize the networking and computing of the BlueField-2's off-path Arm. Michalowicz et al.~\cite{battle_bf3_hoti} compare BlueField-2 and BlueField-3 mainly regarding the off-path Arm's computing ability. In contrast, we focus on characterizing BlueField-3's DPA and Arm from an architectural perspective.

\noindent{\bf SmartNIC offloading.}
Offloading host workloads to SmartNICs has recently attracted significant attention in both academia and industry. Many prior works~\cite{clicknp, smartds, azure_accelnet, tonic, panic, flowblaze, floem, mlweaving_vldb19, e3, lambda_nic, fairnic, fpganic_atc22, shuhai_fccm20, rosebud_asplos23, dfi_sigmod22, alnico_atc22, consensus_nsdi16, hash_join_fpt21, tcp_offload_bf2, enso, hal_isca24, dpu_direct_tc24, linefs, ndp, p4sgd_tpds23, dbnetwork_vldb23, linefs_sosp21, nanopu, k_dist_ml_2, dt_iiswc23, dagger_asplos21, dds_vldb24,lognic} offload host tasks to FPGA-based or SoC-based SmartNICs. PANIC~\cite{panic} addresses the performance isolation and fairness problems under the multi-tenant environment. FlowBlaze~\cite{flowblaze} enables stateful network packet processing on FPGAs. Xenic~\cite{xenic} offloads distributed transactions to SmartNIC. These works leverage SmartNIC to alleviate host CPU pressure but do not provide a comprehensive study on DPA-enhanced SmartNIC.
\vspace{-1ex}
\section{Conclusion}
Typical multicore SmartNICs such as BlueFiled-2 are only capable of processing control-plane tasks with their embedded processors, which cannot directly process network traffic from cloud applications that evolve rapidly over time. Therefore, SmartNIC-related research calls for a programmable datapath accelerator (DPA) to process network traffic at line rate. However, no existing work has unveiled the performance characteristics of DPA. 
We present the first architectural characterization of the latest DPA-enhanced BlueField-3 SmartNIC. We identify that DPA has three unique architectural characteristics that unleash the potential of DPA. We propose three pioneering guidelines for programmers to fully unleash the potential of DPA. To demonstrate the effectiveness of our approach, we conduct detailed case studies regarding each guideline. 
 
\noindent{\bf Acknowledgement. }
\sideword{R2}\zheng{We thank the anonymous ICNP reviewers and our shepherd Sebastiano Miano for their detailed feedback. The work is supported by the following grants: the National Science and Technology Major Project (NO.2022ZD0119301), the National Natural Science Foundation of China (Grant No. 62472384), a research grant from Alibaba Group through Alibaba Innovative Research (AIR) Program, Starry Night Science Fund of Zhejiang University Shanghai Institute for Advanced Study (SN-ZJU-SIAS-0010). Yin Zhang and Zeke Wang are the corresponding authors.}

\bibliographystyle{IEEEtran-Reference-Format}
\bibliography{content/ref}

\end{sloppypar}

\end{document}